\documentclass[aps,prb,twocolumn,showpacs,superscriptaddress,groupedaddress,floatfix,longbibliography]{revtex4-2}
\usepackage{color}
\usepackage{mathtools}
\usepackage{latexsym}
\usepackage{graphicx}
\usepackage{float}
\usepackage{dcolumn}
\usepackage{bm}
\usepackage{amssymb}
\usepackage{amsmath}
\usepackage{mathtools}
\usepackage[space]{grffile}
\usepackage[dvipsnames]{xcolor}
\usepackage[caption=false]{subfig}
\usepackage{ulem}
\usepackage[bookmarks=true,colorlinks,citecolor=RoyalBlue]{hyperref}

\usepackage[colorinlistoftodos]{todonotes}

\definecolor{mygreen}{rgb}{0.25,0.5,0.25}

\begin{document}

\title{Floquet topological systems with flat bands: Edge modes, Berry curvature, and Orbital magnetization}
\author{Ceren B. Dag$^{1,2}$} 
\author{Aditi Mitra$^{3}$} 
\affiliation{$^{1}$ ITAMP, Harvard-Smithsonian Center for Astrophysics, Cambridge, Massachusetts, 02138, USA\\$^{2}$ Department of Physics, Harvard University, 17 Oxford Street Cambridge, MA 02138, USA\\$^{3}$
  Center for Quantum
  Phenomena, Department of Physics, New York University, 726 Broadway,
  New York, NY, 10003, USA}

\begin{abstract}
  Results are presented for Floquet systems in two spatial dimensions where the Floquet driving breaks an effective time reversal symmetry. The driving protocol also induces flat bands that correspond to anomalous Floquet phases where the Chern number is zero and yet chiral edge modes exist. Analytic expressions for the
  edge modes, Berry curvature, and the orbital magnetization are derived for the flat bands. Results are also presented for the static Haldane model for parameters when the bands are flat. Floquet driving of the same model is shown to give rise to Chern insulators as well as anomalous Floquet phases. The orbital magnetization for these different topological phases are presented and are found to be enhanced at half filling by the broken particle-hole symmetry of the Haldane model. 
\end{abstract}

\maketitle

\section{Introduction} \label{intro}
Floquet driving can give rise to new topological phases
that have no analog in static systems \cite{Kitagawa10,Zoller11,Rudner13,Carpentier15,Roy16,Else16b,Kyser-I,Kyser-II,Potter16,Po16,Loss17,Potter17,Morimoto17,Roy17,Roy17a,Fidk19,Yates19,Cirac20,PhysRevB.100.041103,Else16a,Khemani16,Khemani16b,Else17,Yao17,Chandran16,Sondhi20, Natsheh21a,Natsheh21b}. The growing activity in this field is in no small part due to the experimental feasibility of Floquet band engineering \cite{Eckardt17, OkaRev, PhysRevResearch.1.023031, RudnerRev, RevModPhys.93.041002}.
A topic that has been more elusive is detecting these new topological phases, where the experimental tools usually  employed are transport \cite{McIver20} and direct exploration of the spectra through angle resolved photoemission spectroscopy (ARPES) \cite{Gedik13}.

Recent studies have shown that Floquet driving that breaks an effective time-reversal symmetry (TRS) can induce a large
orbital magnetization, which is a linear response of the Floquet system to a small perturbing magnetic field \cite{Mitra22a}.
In the absence of Floquet driving, and in thermal equilibrium, orbital magnetization is well studied theoretically \cite{Ceresoli05,Ceresoli06,Shi07}, and there have also been some experiments that directly or indirectly measure the orbital magnetization \cite{Moriya_2020,PhysRevB.102.121406}. Two dimensional (2D) Floquet systems  that break TRS can also show anomalous phases where the Chern number of the bands are zero, and yet chiral edge modes exist \cite{Rudner13}. The latter gives rise to a quantized orbital magnetization when the bulk
states are localized by spatial disorder \cite{Titum16,Nathan17, Nathan21}. Even in the absence of disorder,
the orbital magnetization of Floquet systems with broken TRS can be significant \cite{Mitra22a}. This observation opens up the possibility of performing  transport and ARPES in the presence of a perturbing magnetic field, the latter making the measurements more sensitive to any Floquet induced topology due to the induced orbital magnetization.

This paper builds on recent results where a general formula for the orbital magnetization for Floquet systems in the absence of spatial disorder,  was derived \cite{Mitra22a}.  In this paper we apply this formula to the cases where the bands are flat, and also to Floquet systems with broken particle-hole symmetry where such a broken symmetry helps to enhance the orbital magnetization at half-filling.

The paper is organized as follows. In Section \ref{model} the models that will be studied are introduced. In Section \ref{Kit}, analytic results for the edge modes, Berry curvature, and the orbital magnetization are presented in the limit of flat bands. In Section \ref{HaldaneF},
results are presented for the orbital magnetization for the static Haldane model \cite{Haldane88}
for the case where the bands are flat \cite{Mudry11}. Following this, the Floquet driven Haldane model is studied, and 
results for the orbital magnetization for Chern insulator phases
and for anomalous Floquet phases are given. 
Finally we present our conclusions in Section \ref{Conclu}.
Intermediate steps in the derivation of analytic expressions are provided in three appendices. 

\begin{figure}
\centering{\includegraphics[width=0.49\textwidth]{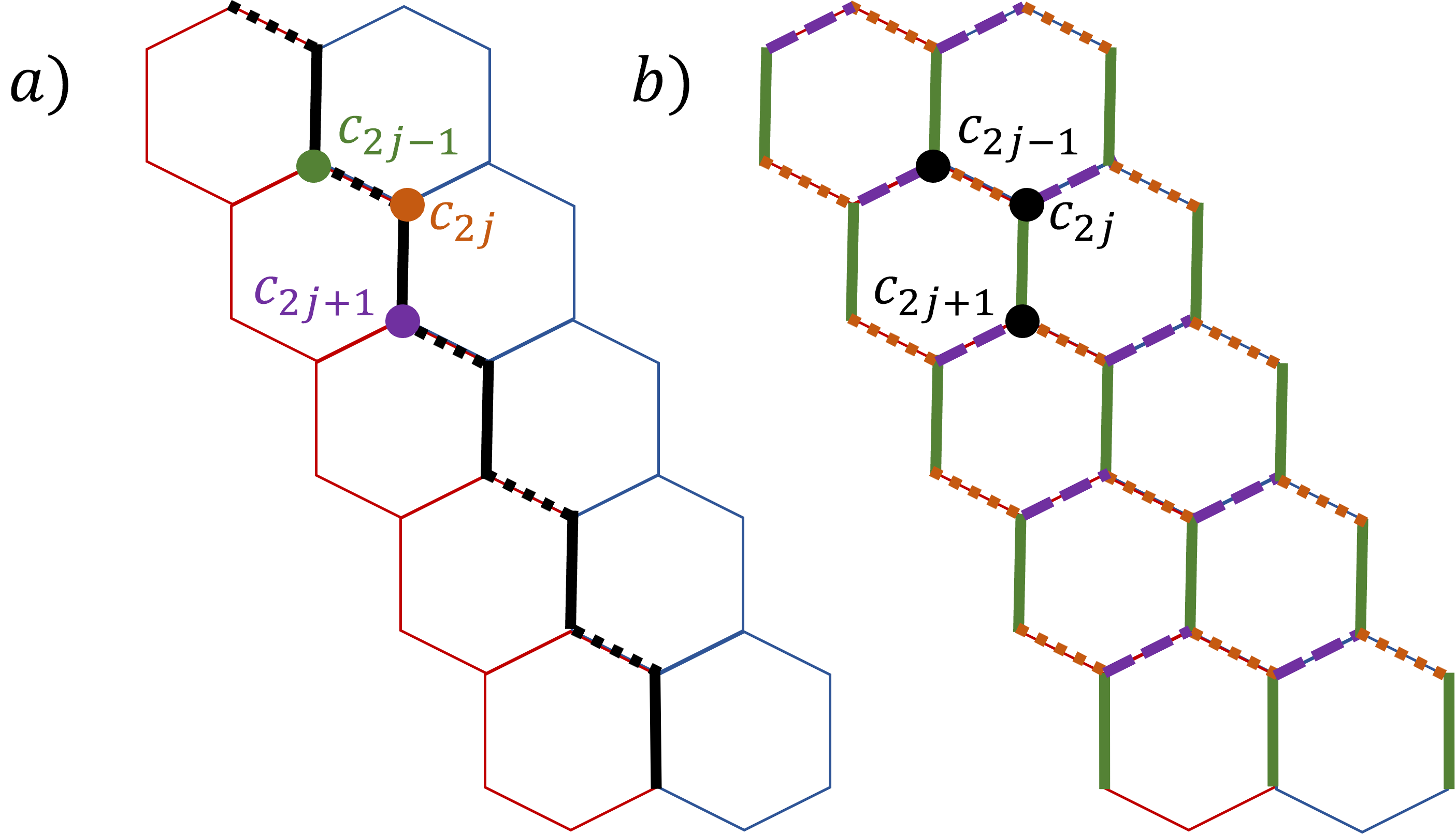}}\hfill 
\caption{(a) Schematic of static graphene with a zigzag boundary. The dotted and solid black lines indicate different coupling strengths on the cylinder, resulting in an effective Su-Schrieffer-Heeger model \cite{SSH79,SSH80}. (b) Schematic of Floquet driven graphene with a zigzag boundary. The first, second and third steps of the Floquet protocol are represented by green solid, purple dashed and orange dotted lines respectively.}
\label{Fig1}
\end{figure}

\section{Models} \label{model}

We study two classic models, one of graphene \cite{NetoRev} and the other of the Haldane model \cite{Haldane88}. We study these two models both under periodic boundary conditions (i.e.,~on a torus), and on a cylinder. The cylindrical geometry has the advantage that it shows edge modes, and is particularly helpful for identifying anomalous Floquet phases where the Chern number does not fully characterize the number of chiral edge modes. For example, one may have bands with zero Chern number, and yet chiral edge modes can exist \cite{Rudner13}.

We employ the Floquet driving protocol of Ref.~\onlinecite{Kitagawa10} where for the case of graphene, within a drive cycle, the three nearest neighbor (n.n.)~hopping parameters are modulated cyclically. When we Floquet drive the Haldane model, the n.n. hopping parameters are similarly modulated, but we keep the next-nearest-neighbor (n.n.n.) hoppings and the flux constant in time. 

On a cylindrical geometry, we take the $x$ direction to be periodic and the $y$ direction to be open, with a total number of $N \in $ even sites. The convention for the distances between n.n. sites is chosen to be
\begin{align}
  \boldsymbol{\delta}_1 =a \left(0,-1\right), \,\,
  \boldsymbol{\delta}_2 = \frac{a}{2}\left(\sqrt{3}, 1\right), \,\, \boldsymbol{\delta}_3 = \frac{a}{2}\left(-\sqrt{3}, 1\right),
\end{align} 
where $a$ is the lattice spacing.

\begin{figure}
\centering{\includegraphics[width=0.49\textwidth]{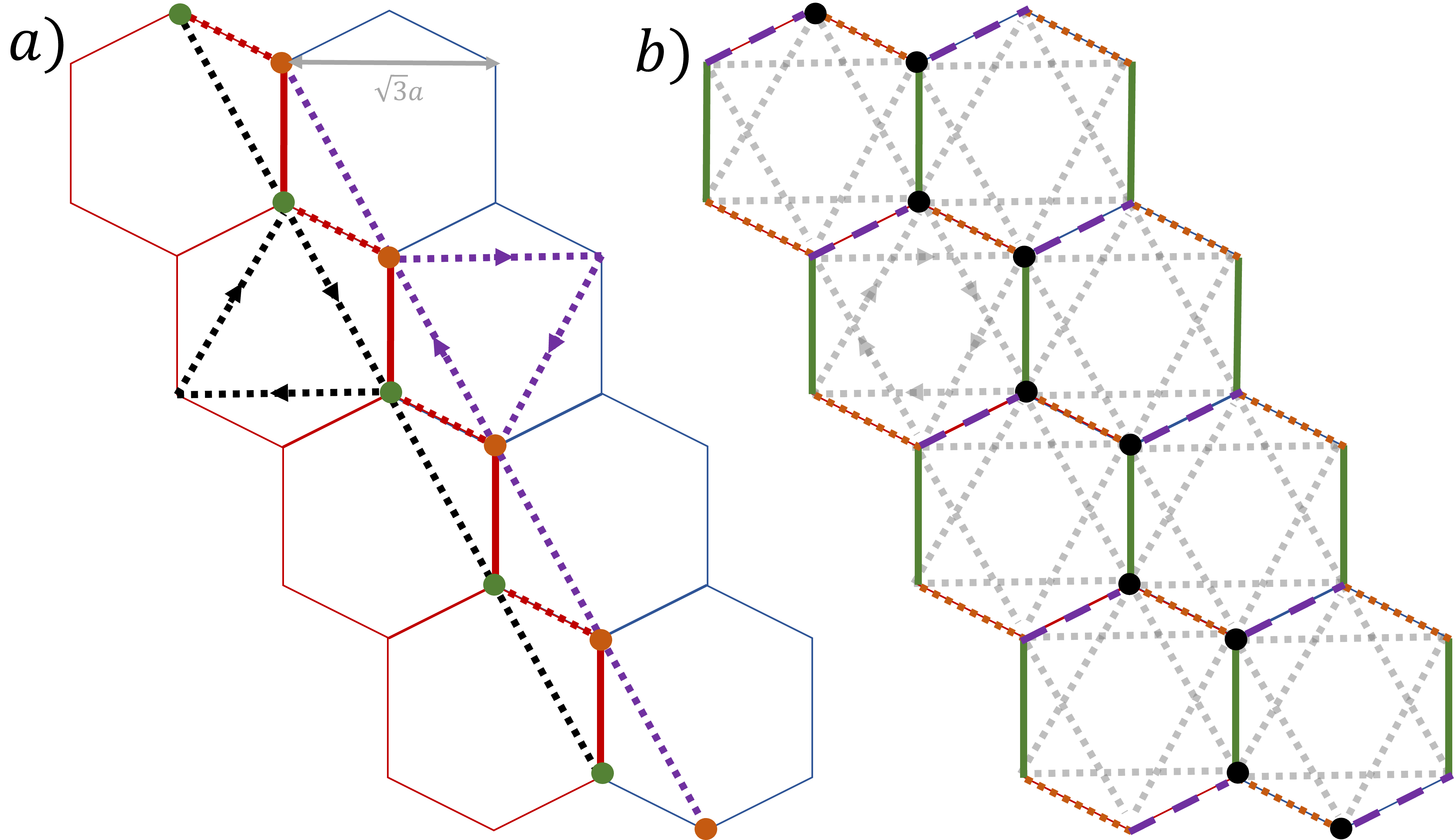}}\hfill 
\caption{(a) Static Haldane model on a cylinder with a zigzag boundary. The black and purple dotted lines connect odd (green) and even (orange) sites, respectively, whose coupling strengths differ on the boundary due to broken TRS (second line of Eq.~\eqref{HaldaneOnCylinder}). Horizontal lines denote n.n.n. couplings, and on the boundary these again contribute differently for even and odd sites (third line of Eq.~\eqref{HaldaneOnCylinder}). The TRS breaking phase $\phi$ is either positive or negative depending on the direction of the arrows. (b) Floquet driven Haldane model on a cylinder where the first, second and third steps of the protocol are represented by green solid, purple dashed and orange dotted lines respectively. The 
n.n.n.~couplings are pictured as gray and indicate that they are not driven but remain fixed at a nonzero value.}
\label{Fig2}
\end{figure}

\begin{figure*}
\subfloat[]{\label{Fig3a}\centering{\includegraphics[width=0.30\textwidth]{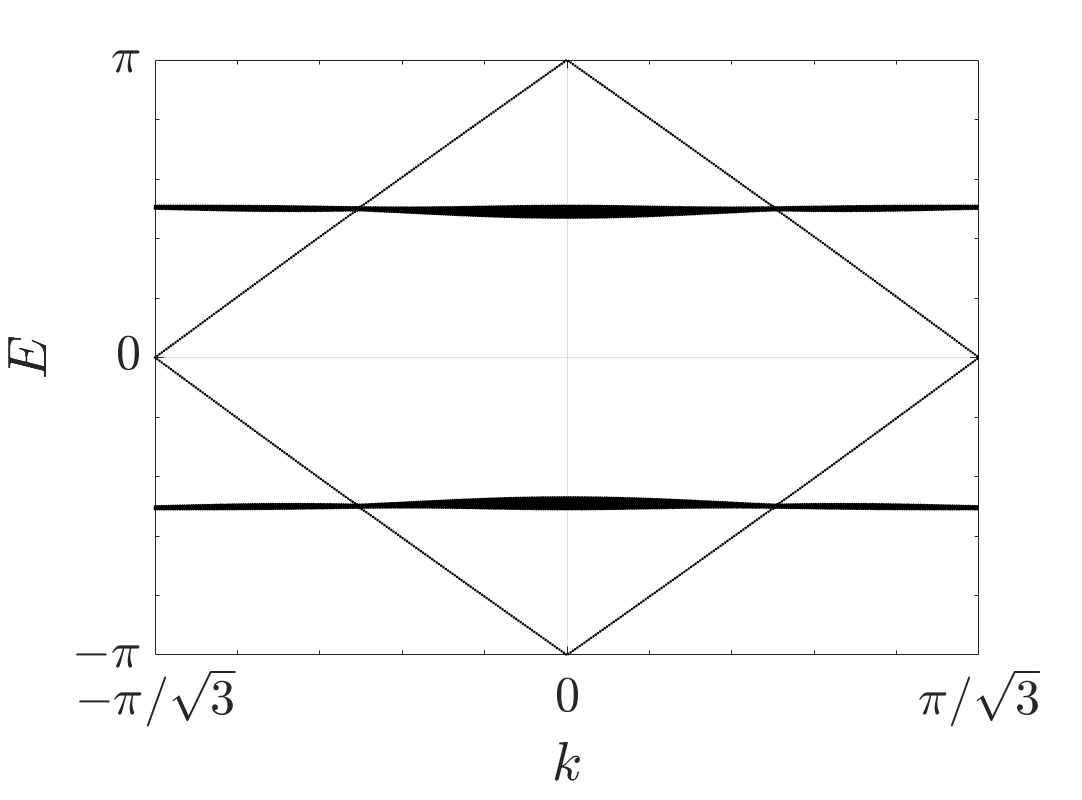}}}\hfill 
\subfloat[]{\label{Fig3b}\centering{\includegraphics[width=0.33\textwidth]{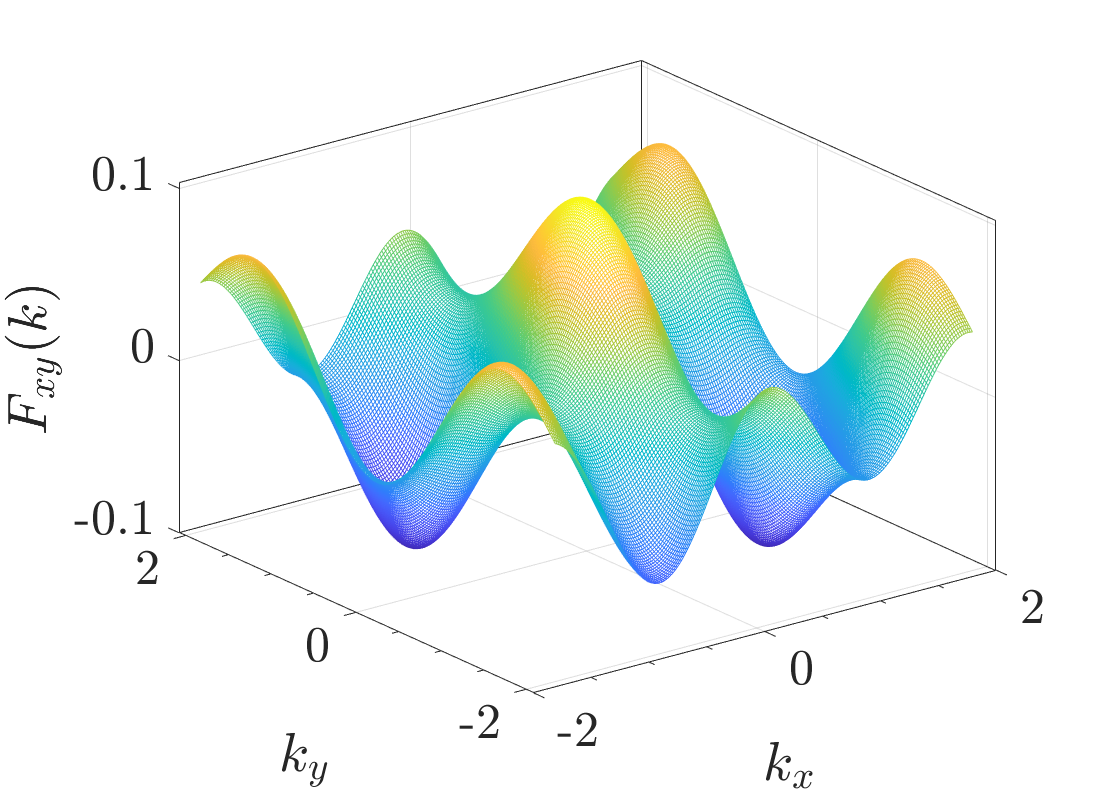}}}\hfill
\subfloat[]{\label{Fig3c}\centering{\includegraphics[width=0.32\textwidth]{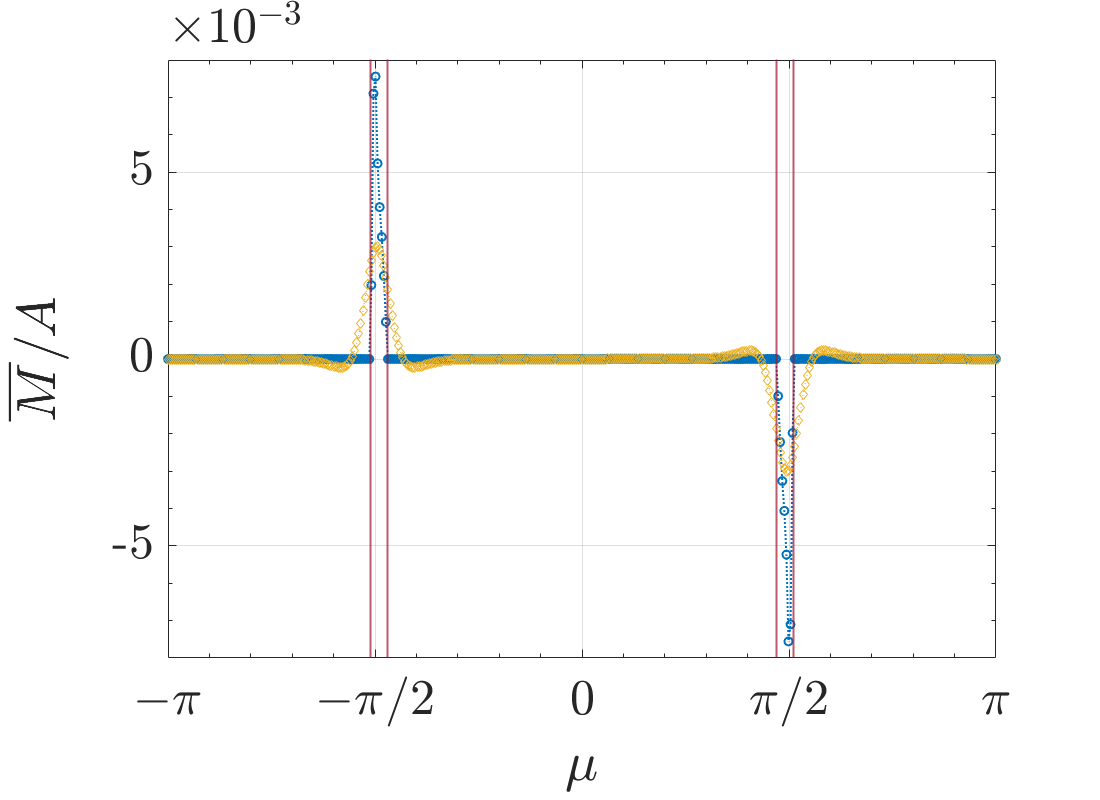}}}\hfill
\caption{(a) Band structure of driven graphene with zigzag boundaries for small detuning parameter $JT/3=0.01\times \pi/2$ from the flat band limit while keeping $\lambda JT/3= \pi/2$ fixed. The bands are not perfectly flat because $JT \neq 0$. (b). Berry curvature for the model on a torus for the same parameters. The integral of the Berry curvature over the BZ vanishes i.e.,~$C=0$. (c). Orbital magnetization per unit area on the torus, in units of $-e/\hbar T$, for
the same parameters with a temperature $\beta^{-1}= 0.05 \times J T/3$ (blue-circles), and a temperature $\beta^{-1}= 0.05 \times \lambda J T/3$ (yellow-diamonds). Vertical red lines denote the band edges.}
\label{Fig3}
\end{figure*}

At each momentum point $k$, and for a zig-zag boundary (see Fig.~\ref{Fig1}), the Hamiltonian is
\begin{widetext}
\begin{eqnarray}
&H(t)= -\sum_{j,k} \biggl[J_1(t)\left(c_{2j,k}^{\dagger}c_{2j+1,k} + h.c.\right)
+\biggl\{c_{2j,k}^{\dagger}c_{2j-1,k} \left(J_2(t)e^{i k a \sqrt{3}/2}+ J_3(t)
e^{-i k a \sqrt{3} /2}\right) + h.c.\biggr\}\biggr].
\end{eqnarray}
The couplings $J_{1,2,3}(t)$ are periodically modulated according to the following protocol \cite{Kitagawa10}
\begin{subequations}\label{Jprofile}
\begin{eqnarray}
&(1)\hspace{5mm}& J_1=\lambda J; \hspace{2mm} J_2=J_3=J \hspace{2mm}\text{for } n T <t\leq n T+ T/3, \\
&(2)\hspace{5mm}& J_2=\lambda J; \hspace{2mm} J_1=J_3=J \hspace{2mm}\text{for } n T+T/3<t\leq n T+2 T/3, \\
&(3)\hspace{5mm}& J_3=\lambda J; \hspace{2mm} J_1=J_2=J \hspace{2mm}\text{for } n T+2 T/3<t\leq n T+ T, 
\end{eqnarray}
\end{subequations}
\end{widetext}
with $T$ being the period. Note that $\lambda=1$ corresponds to the static case. In what follows we will express all energies
in units of $T^{-1}$. Denoting the quasi-energies by $\epsilon$, we will choose the Floquet Brillouin zone (FBZ) to be between $\epsilon \in \left[-\pi,\pi\right]$. The driving scheme outlined above goes by the name of quantum walks \cite{Kitagawa11}, and is an alternative to shining light on graphene \cite{Gedik13,McIver20} or periodically modulating an optical lattice \cite{Esslinger14,goldman_cooper_dalibard_2017}. In fact, it has recently been shown that the above driving protocol can be implemented in Rydberg gases \cite{2021arXiv210111412K}.

For the protocol of Eq.~\eqref{Jprofile}, the Floquet unitary is
\begin{align}\label{U}
    U = U_3 U_2 U_1,
\end{align}
where $U_n = e^{-i H_n T/3}$ and 
\begin{align*}
    H_n = H\biggl(m T + (n-1) T/3<t<m T+ n T/3\biggr),
\end{align*}
with $n=1,2,3$ and $m$ is an integer.

The second model we study is the Haldane model with the
n.n.~hoppings modulated according to Eq.~\eqref{Jprofile}. On
a cylindrical geometry (see Fig.~\ref{Fig2}) the model is
\begin{widetext}
\begin{eqnarray}\label{HalF}
H(t) &=& \sum_{j,k}\biggl[ J_1(t)\left(c_{2j,k}^{\dagger}c_{2j+1,k} + h.c.\right) +\biggl\{c_{2j,k}^{\dagger}c_{2j-1,k} \left(J_2(t)e^{ik a \sqrt{3}/2}+J_3(t)e^{-ik a \sqrt{3}/2}\right) + h.c.\biggr\}, \notag \\
&+& t_2 \biggl\{(e^{-i k a \sqrt{3} /2} e^{-i \phi} + e^{i k a \sqrt{3}/2} e^{i \phi}) c_{2j-1,k}^{\dagger}c_{2j+1,k}+h.c.\biggr\} +
t_2 \biggl\{(e^{-i k a \sqrt{3}/2} e^{i \phi} + e^{i k a \sqrt{3}/2} e^{-i \phi}) c_{2j,k}^{\dagger}c_{2j+2,k}+h.c.\biggr\}
\notag \\
&+&t_2 \biggl\{(e^{-ika \sqrt{3}} e^{i \phi} + e^{ik a \sqrt{3}} e^{-i \phi}) c_{2j-1,k}^{\dagger}c_{2j-1,k} +(e^{-i ka \sqrt{3}} e^{-i \phi} 
+ e^{i k a \sqrt{3}} e^{i \phi}) c_{2j,k}^{\dagger}c_{2j,k} \biggr\}\biggr].\label{HaldaneOnCylinder}
\end{eqnarray}
\end{widetext}
The first line represents the usual n.n.~hopping terms also encountered in graphene. The second and third lines represent the n.n.n.~hopping with an amplitude $t_2$ and a TRS breaking  flux $\phi$. %An inversion symmetry breaking mass $M$ is also added for generality. However in this paper we present results for $M=0$.

We are interested in studying the orbital magnetization for these models. In general when TRS is broken, a perturbing magnetic field can induce an orbital magnetization. This linear response behavior has been extensively studied for systems in thermal equilibrium \cite{Ceresoli05,Ceresoli06,Shi07}, where the orbital magnetization ${\bf M}$ is simply proportional to the rate at which the free-energy $F$ changes when a perturbing magnetic field ${\bf B}$ is applied, i.e $F=F(B=0)+{\bf M}\cdot{\bf B}$. In the Floquet context, it was shown that the orbital magnetization of a Floquet eigenstate, averaged over one drive cycle, is equal to the rate of change of the quasi-energy due to the applied magnetic field \cite{Nathan17}. Thus the total orbital magnetization averaged over one drive cycle is related to the $\mathcal{O}(B)$ shift of the quasi-energy of each Floquet eigenstate, but weighted with the occupation of the Floquet eigenstate. At $\mathcal{O}(B)$, the occupation of the Floquet eigenstates does not change as at $O(B)$ only virtual processes can occur, and any change in the occupation requires real inelastic process. With this as the starting point, and under the assumption that the occupation of the Floquet eigenstates is given by the Fermi-Dirac distribution function, the orbital magnetization averaged over one drive cycle is \cite{Mitra22a}
\begin{align}
&\overline{M} =-\frac{e}{2\hbar}{\rm Im}\biggl[\sum_{n{\bf k}}f_{n {\bf k}}\nonumber\\
&\times    \overline{\langle \partial_{{\bf k}} \phi_{n{\bf k}}(t) |\left(\epsilon_{n {\bf k}}+H_F-2\mu\right)\times |\partial_{\bf k}\phi_{n {\bf k}}(t)\rangle}\nonumber\\
&-f'_{n {\bf k}}\left(\epsilon_{n {\bf k}}-\mu\right)\nonumber\\
&\times  \overline{\langle \partial_{\bf k}\phi_{n{\bf k}}(t) | \left(\epsilon_{n {\bf k}}- H_F\right)  \times |\partial_{\bf k}\phi_{n {\bf k}}(t)\rangle}\biggr]. \label{Meq1}
\end{align}
Above $f_{n{\bf k}}$ represents the Fermi-Dirac distribution at a temperature $\beta^{-1}$ and at a chemical potential $\mu$.
$f'_{n, {\bf k}}$ denotes derivative of the Fermi function with respect to the energy. The combination $f'(x) x$ ensures that the second term only contributes at non-zero temperatures.
$\epsilon_{n, {\bf k}}$ is the quasi-energy labeled by the band $n$ and the quasi-momentum ${\bf k}$. 
$|\phi_{n {\bf k}}(t)\rangle$ is the corresponding Floquet quasimode, while  
$H_F = H(t) - i\partial_t$ or equivalently $H_F T = i\ln{U}$
is the Floquet Hamiltonian. 
In addition, the notation $\overline{O}$ denotes time-average over one drive cycle i.e.,~$\overline{O}= \int_t^{t+T}dt' O(t')/T$. 
In the absence of a drive, the above expression reduces to the orbital magnetization of static systems, excluding corrections to the orbital magnetization coming from changes to the entropy \cite{Ceresoli05,Ceresoli06,Shi07}. 

The above form of the orbital magnetization is also natural if one notes that the orbital magnetization is proportional to the
average of ${\bf r} \times {\bf v}= {\bf r} \times i\left[H,{\bf r}\right]$ in each eigenstate, with ${\bf r}$ being the position operator, and ${\bf v}={\dot{\bf r}}$ being the velocity operator. Noting that $\bf{r}$ acts as $\partial_{\bf k}$ on momentum eigenstates, the above expression for the orbital magnetization with its dependence on the Berry curvature, the quasi-energy and the Floquet Hamiltonian, naturally emerges \cite{Ceresoli05,Ceresoli06}.

Of course, it is not guaranteed that the Floquet eigenstates will be occupied according to the Fermi-Dirac distribution function. In fact, when the system is coupled to an ideal reservoir in thermal equilibrium, it is only in the limit of high frequency driving that the Floquet eigenstates acquire a thermal equilibrium occupation. For general driving protocols, the occupations can be quite complicated by depending upon the details of the reservoir and the system-reservoir coupling \cite{PhysRevB.90.195429,Dehghani15a,PhysRevB.91.235133,PhysRevX.5.041050,PhysRevE.91.030101,Dehghani16b,Shirai_2016,PhysRevA.98.033601}. Another natural choice of the occupation probability is a quench occupation probability which is simply given by the mod-square of the overlap of the Floquet eigenstates with the eigenstates before the drive was switched on, and  weighted by the occupation of the undriven states. Such a quench distribution was already considered in the study of the orbital magnetization in Ref.~\cite{Mitra22a} where it was shown that a quench occupation can be more sensitive to van-Hove singularities, and can even take larger values than for a thermal occupation of the bands. In addition, if the Floquet eigenstates are too different from the eigenstates of the undriven system, the effective temperature can be very high, and that can in turn smooth out many features. Hence, for the purpose of highlighting the key physics and still keeping the discussion simple, we assume that the occupation probabilities are given by a Fermi-Dirac distribution function.

For a two band model, with the Floquet bands labeled by $n=u,d$, the above formula simplifies to
\begin{align}
  \overline{M}
  & =-\frac{e}{2\hbar}{\rm Im}\sum_{\boldsymbol{k}}\biggl[\left(f_{d\boldsymbol{k}}-f_{u\boldsymbol{k}}\right)\left(\epsilon_{d\boldsymbol{k}}+\epsilon_{u\boldsymbol{k}}-2\mu\right)\overline{F_{xy}(\boldsymbol{k},t)}\nonumber\\
&-(\epsilon_{d\boldsymbol{k}}-\epsilon_{u\boldsymbol{k}})\overline{F_{xy}(\boldsymbol{k},t)}\sum_{n=d,u}f'_{n\boldsymbol{k}}(\epsilon_{n\boldsymbol{k}}-\mu)\biggr].\label{Meq2}
\end{align}
Above $\overline{F_{xy}(k,t)}$ is the Berry curvature averaged over one drive cycle. Since in this paper we will be working with a Floquet unitary $U$ that generates stroboscopic time-evolution rather than a time-dependent Hamiltonian, and therefore we will not have information on the micro-motion within a drive cycle, we will replace $\overline{F_{xy}(k,t)}$
in the above equation by $F_{xy}(k)$ where $F_{xy}$
is the curvature obtained from the eigenmodes of $U$.

It is interesting to note that due to the term $\epsilon_{d\boldsymbol{k}}+\epsilon_{u\boldsymbol{k}}$ in Eq.~\eqref{Meq2}, when the bands break particle-hole symmetry (i.e., $\epsilon_{d\boldsymbol{k}}+\epsilon_{u\boldsymbol{k}}\neq 0$),
the orbital magnetization is  enhanced relative to the case of particle-hole symmetric systems. This will be apparent when we study the Haldane model for generic values of the flux $\phi$.

Before we present our results, note that while the orbital magnetization for the disorder-free system depends on the Berry curvature, the values taken by it will be non-universal as the integral involves the product of the Berry curvature and the quasi-energy. When states are localized due to spatial disorder, Ref.~\cite{Nathan17} showed that the orbital magnetization is quantized and equal to the 3D winding number introduced in \cite{Rudner13}. For our case, the bulk states are conducting which causes the orbital magnetization to acquire both topological and non-topological contributions that are difficult to separate.

\section{Analytic expressions in the flat band limit}\label{Kit}

In this section we will obtain analytic expressions for flat bands. 
Perfectly flat bands are obtained on taking the following limits
\begin{equation}\label{limit}
J T \rightarrow 0; \, \lambda \rightarrow \infty; \,  
\lambda J T/3 = \pi/2,
\end{equation}
with the unitaries taking the simple form
\begin{subequations}\label{Ulim}
\begin{align}
  &U_1 = e^{i\pi/2\left(c_{2j,k}^{\dagger}c_{2j+1,k} + h.c.\right)},\\
  & U_2 =  e^{i \pi/2\left(c_{2j,k}^{\dagger}c_{2j-1,k} e^{i k a \sqrt{3}/2} + h.c.\right)},\\
  & U_3 = e^{i \pi/2 \left(c_{2j,k}^{\dagger}c_{2j-1,k}e^{-i k a \sqrt{3}/2} + h.c.\right)}.
\end{align}
\end{subequations}
In this limit, the effect of periodic driving can be visualized in terms of the tunneling of the fermions between n.n.~sites at each application of $U_i$. Depending on the order of the unitaries, e.g.,~either $U_3 U_2 U_1$ or $U_1 U_2 U_3$, one can set the direction of the chirality. We are interested in the stroboscopic time-evolution corresponding to 
\begin{align}
  \left[U\right]^{\dagger}c_{j,k} \left[U\right]= \left[U_3 U_2 U_1\right]^{\dagger}c_{j,k}\left[U_3 U_2 U_1\right]=\sum_l\tilde{U}_{j,l}c_{l,k}. \notag
\end{align}
The matrix $\tilde{U}$ is a unitary matrix. The above linear relation between fermion operators after a stroboscopic time step, and before, is due to the free fermion nature of the problem. 
For the unitaries corresponding to Eq.~\eqref{Ulim}, $\tilde{U}$ 
takes the following form for a 8-site system
(see Appendix \ref{appA} for details)
\begin{widetext}
\begin{eqnarray}
\tilde{U} = \left( \begin{array}{c c c c c c c c}
-e^{i k a \sqrt{3}} & 0 & 0 & 0 & 0 & 0 & 0 & 0 \\
0 & 0 &  -ie^{ -ika \sqrt{3}} & 0 & 0 & 0 & 0 & 0  \\
0 &  -ie^{ika \sqrt{3}} & 0 & 0 & 0 & 0 & 0 & 0  \\
0 & 0 & 0 & 0 &  -ie^{-ika \sqrt{3}} & 0 & 0 & 0  \\
0 & 0 & 0 &  -ie^{ika \sqrt{3}} & 0 & 0 & 0 & 0  \\
0 & 0 & 0 & 0 & 0 & 0 & -ie^{-ik a \sqrt{3}} & 0  \\
0 & 0 & 0 & 0 & 0 & -ie^{ika \sqrt{3}} & 0 & 0  \\
0 & 0 & 0 & 0 & 0 & 0 & 0 & -e^{ -i k a \sqrt{3}}  \\
\end{array} \right), \label{generalU}
\end{eqnarray}
\end{widetext}
where the basis explicitly is $(c_{1,k},c_{2,k},\dots,c_{8,k})$.
The quasi-energies are obtained from $i\ln{\tilde{U}}$.
(For an analytic expression for $U$ itself in terms of the number operators, see Appendix~\ref{appC}). Because $\tilde{U}$ is block diagonal, one can easily generalize it to the thermodynamic limit. In other words, as the number of sites is increased, and stays even, the second and third rows keep repeating, whereas the first and last rows incorporate the edge modes.

The first and last rows of $\tilde{U}$ clearly show that there are chiral edge modes with a dispersion $ka\sqrt{3}$ on one end of the cylinder, and $-ka\sqrt{3}$ on the other end. Moreover close to $k=0$, these edge modes cross the Floquet zone boundaries $\epsilon=\pm \pi$, while at 
the edges of the momentum Brillouin zone, $ka =\pm \pi/\sqrt{3}$, the edge modes cross the center of the FBZ $\epsilon=0$. Thus we recover
the edge modes of a anomolous phase where edges modes of the same chirality exist on either sides of the bulk bands.
There are also bulk states with all the bulk states having the same two quasi-energies $\epsilon=\pm \pi/2$. Thus the bulk bands are perfectly flat, and highly degenerate. Although we work with zigzag boundaries to derive the flat band limit, armchair boundaries can also lead to flat bulk bands and chiral edge modes \cite{Kitagawa10}.

A plot of the spectrum is shown in Fig.~\ref{Fig3a}. The parameters are slightly detuned from the perfectly flat band limit. The results agree with the analytic expressions. The chiral edge modes are clearly visible,
and so are the two bands, whose centers are located at $\epsilon=\pm \pi/2$.

We now turn to the discussion of the Berry curvature and the orbital magnetization of the flat bands. For this we now impose periodic boundary conditions, i.e.,~consider a torus geometry. 
In the limit corresponding to Eq.~\eqref{limit}, the Floquet unitary on a torus takes a simple form
\begin{eqnarray}
U_F &=& \prod_n e^{-iH_n T/3} = -i \prod_n \left[\cos(\boldsymbol{k}\cdot \boldsymbol{\delta}_n) \sigma_x - \sin(\boldsymbol{k}\cdot \boldsymbol{\delta}_n) \sigma_y \right] \notag \\
&=& \left( \begin{array}{c c}
0 & -ie^{i \boldsymbol{k}\cdot \boldsymbol{\bar{\delta}}} \\
-ie^{-i\boldsymbol{k}\cdot \boldsymbol{\bar{\delta}}} & 0  
\end{array} \right) \label{UF1}.
\end{eqnarray}
Above $\boldsymbol{\delta}_1-\boldsymbol{\delta}_2+\boldsymbol{\delta}_3=(-\sqrt{3},-1)\equiv \boldsymbol{\bar{\delta}}$.
Eq.~\eqref{UF1} implies that $U_F^2=-I$. Hence two periods of driving guarantees localization in the bulk. In other words, a fermion that started tunneling in the bulk will always stay in the bulk, leading to localized bulk states. However this localization is absent for the eigenstates of $U_F$.

On diagonalizing $U_F$, the quasi-energies are $\epsilon=\pm \pi/2$, as expected, while the quasimodes are
\begin{eqnarray}
|{\psi^{\pm}}\rangle&=& \frac{1}{\sqrt{2}}\left( \begin{array}{c}
e^{i \boldsymbol{k}\cdot \boldsymbol{\bar{\delta}}} \\
\pm 1   
\end{array} \right).
\end{eqnarray}
It is straightforward to see that the Berry curvature is zero because we have a spinor that lies completely in the $x-y$ plane.  Only 
when we are slightly detuned from the flat band limit, the curvature becomes nonzero, but small, as seen in Fig.~\ref{Fig3b}. From
Eq.~\eqref{Meq2}, zero Berry curvature implies that the orbital magnetization is also zero.

To obtain a non-zero result for the Berry curvature, we will now perturb around the limit in
Eq.~\eqref{limit}. There are two ways to perturb around the flat band limit. One is to take $JT$ to be non-zero but small $JT \ll 1$ while keeping $\lambda JT/3=\pi/2$ fixed. A second way involves keeping $JT=0$, but tuning the parameters as follows
\begin{equation}\label{limit2}
J T \rightarrow 0;\, \lambda \rightarrow \infty;\, 
\lambda J T/3 = \pi/2 + \xi; \,\xi \ll 1.
\end{equation}
We show the results of the former approach numerically in Figs.~\ref{Fig3}, while analytically treating the  the perturbation in Eq.~\eqref{limit2}. Hence, in what follows we will perform a systematic expansion in powers of $\xi$. Both perturbation methods result in the same qualitative structure for the orbital magnetization.

Denoting the two bands as $n=\pm$, we find (Appendix \ref{appB})
\begin{align}
    \epsilon_{n \boldsymbol{k}} &=\pm \frac{\pi}{2} \pm  
    \xi \epsilon_1(\boldsymbol{k}) \pm O(\xi^2),
\end{align}
where
\begin{align}\label{eps1}
    \epsilon_1(\boldsymbol{k}) = \cos(\sqrt{3}k_x)+2\cos\left(\frac{3k_y}{2}\right)
    \cos\left(\frac{\sqrt{3}k_x}{2}\right).
\end{align}
 In addition we find that the Berry curvature is (see Appendix~\ref{appB} for details) 
\begin{widetext}
\begin{eqnarray}\label{Fxy0}
F_{xy}(\boldsymbol{k}) &=& -\frac{\sqrt{3}}{2} \xi \left[\cos(\sqrt{3}k_x)+2\cos\left(\frac{\sqrt{3}k_x}{2}-\frac{3k_y}{2}\right)+\cos\left(\frac{\sqrt{3}k_x}{2}+\frac{3k_y}{2}\right) \right] + O(\xi^3).
\end{eqnarray}
\end{widetext}
On integrating the Berry curvature over the momentum Brillouin zone, we indeed obtain $C=0$. Despite the Chern number vanishing,
a non-zero Berry curvature will be important for obtaining a non-zero orbital magnetization. We discuss this further below.

For computing the orbital magnetization 
we need the difference in occupations of the two bands. We find this to be (see Appendix~\ref{appB})
\begin{widetext}
\begin{eqnarray}\label{fdiff}
f_{d\boldsymbol{k}} - f_{u\boldsymbol{k}} 
&=& \frac{\sinh\left(\beta\pi/2\right)}{\cosh\left(\beta\pi/2\right)+\cosh\left(\beta\mu\right)} +\xi  \frac{\left[1+\cosh\left(\beta\pi/2\right)
\cosh\left(\beta\mu\right) \right]}{\left[\cosh\left(\beta\pi/2\right)
+\cosh\left(\beta\mu\right)\right]^2} \beta\epsilon_1(\boldsymbol{k}) + O(\xi^2).
\end{eqnarray}
We also note that $\epsilon_{d\boldsymbol{k}} + \epsilon_{u\boldsymbol{k}} =0$ based on assumptions of Eq.~\eqref{limit2}.

We also need the following expression to compute the part of the orbital magnetization that contributes only at non-zero temperature (Appendix~\ref{appB})
\begin{align}
&-\sum_{n=d,u}f'_{n\boldsymbol{k}}\beta(\epsilon_{n\boldsymbol{k}}-\mu) =2  e^{2\gamma^+} \left(\frac{e^{-2\beta  \mu }\gamma^- }{\left(1+e^{2\gamma^-}\right)^2}-\frac{\gamma^+ }{\left(1+e^{2\gamma^+}\right)^2}\right)\nonumber\\
&+\frac{1}{8} \beta  \epsilon_1(\boldsymbol{k}) \xi \bigl[\bigl(2-4\gamma^- \tanh \gamma^-\bigr) \text{sech}^2 \gamma^- -\bigl(2- 4\gamma^+ \tanh \gamma^+ \bigr) \text{sech}^2\gamma^+\bigr]+O\left(\xi^2\right), \label{Eq19}
\end{align}
where $\gamma^{\pm} = \beta  (\pi \pm 2 \mu )/4$ is a dimensionless quantity. Thus the orbital magnetization per unit area is found to be
\begin{align}
\overline{M}/A &= \frac{e\mu}{\hbar} \xi  \frac{\left[1+\cosh\left(\beta\pi/2\right)
\cosh\left(\beta\mu\right) \right]}{\left[\cosh\left(\beta\pi/2\right)
+\cosh\left(\beta\mu\right)\right]^2} \int \frac{d\bm{k}}{(2\pi)^2} F_{xy}(\boldsymbol{k}) \,\, \beta \epsilon_1(\boldsymbol{k})\nonumber\\
&-\frac{e}{16\hbar} \xi  \bigg [  \biggl\{ 4(\mu-\pi) + 4 \pi \gamma^- \tanh \gamma^- \biggr\} \text{sech}^2\gamma^- + \biggl\{ 4(\mu+\pi) -4\pi \gamma^+ \tanh \gamma^+\biggr\} \text{sech}^2\gamma^+\bigg] \int\frac{d\bm{k}}{(2\pi)^2} F_{xy}(\boldsymbol{k}) \,\, \beta \epsilon_1(\boldsymbol{k}). 
\end{align}
Substituting for $F_{xy}(\boldsymbol{k})$ from Eq.~\eqref{Fxy0}, and $\epsilon_1(\boldsymbol{k})$
from Eq.~\eqref{eps1}, we obtain
\begin{align}
\overline{M}/A 
&= \xi^2 \frac{2}{3} \frac{e}{\hbar} \bigg[\mu\beta\frac{\left[1+\cosh\left(\beta\pi/2\right)
\cosh\left(\beta\mu\right) \right]}{\left[\cosh\left(\beta\pi/2\right)
+\cosh\left(\beta\mu\right)\right]^2}- \frac{\beta }{16}  \biggl\{ 4(\mu-\pi) + 4\pi \gamma^- \tanh \gamma^- \biggr\} \text{sech}^2\gamma^-  \nonumber\\
&-\frac{\beta }{16} \biggl\{ 4(\mu+\pi) - 4\pi \gamma^+ \tanh \gamma^+ \biggr\} \text{sech}^2\gamma^+\bigg]
+    \mathcal{O}(\xi^3).\label{Mex}
\end{align}
\end{widetext}

\begin{figure*}
\subfloat[]{\label{Fig4a}\centering{\includegraphics[width=0.45\textwidth]{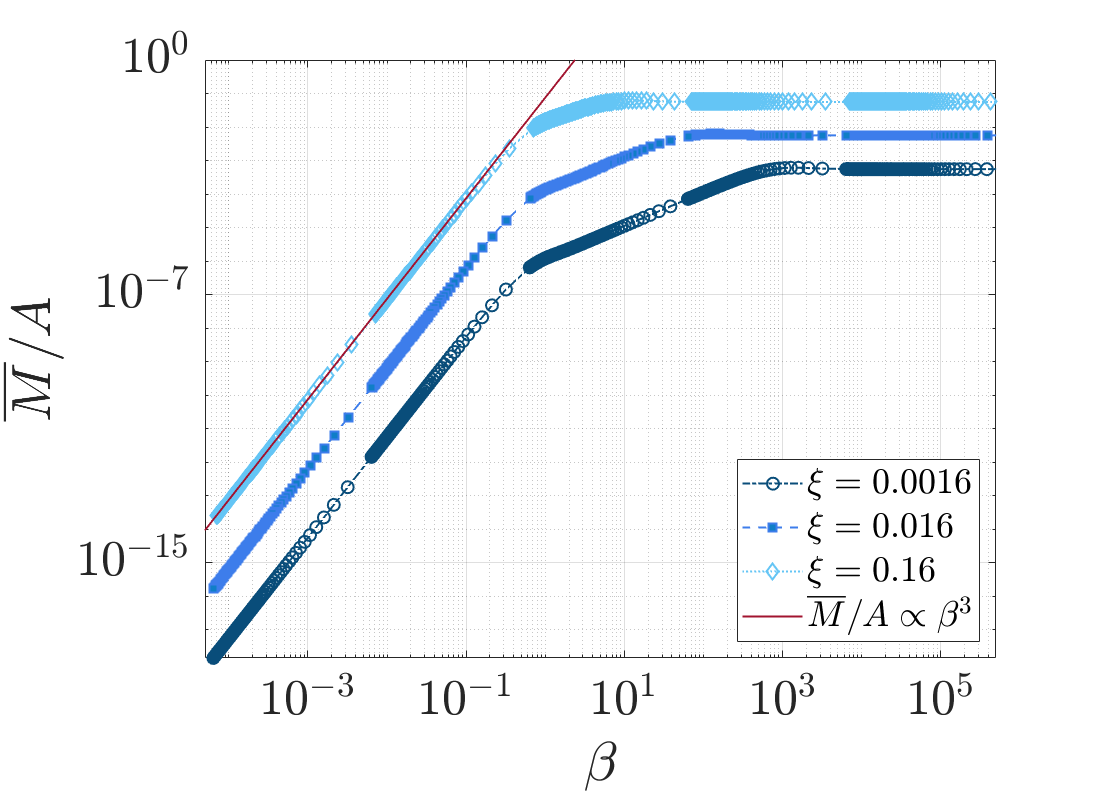}}}
\subfloat[]{\label{Fig4b}\centering{\includegraphics[width=0.45\textwidth]{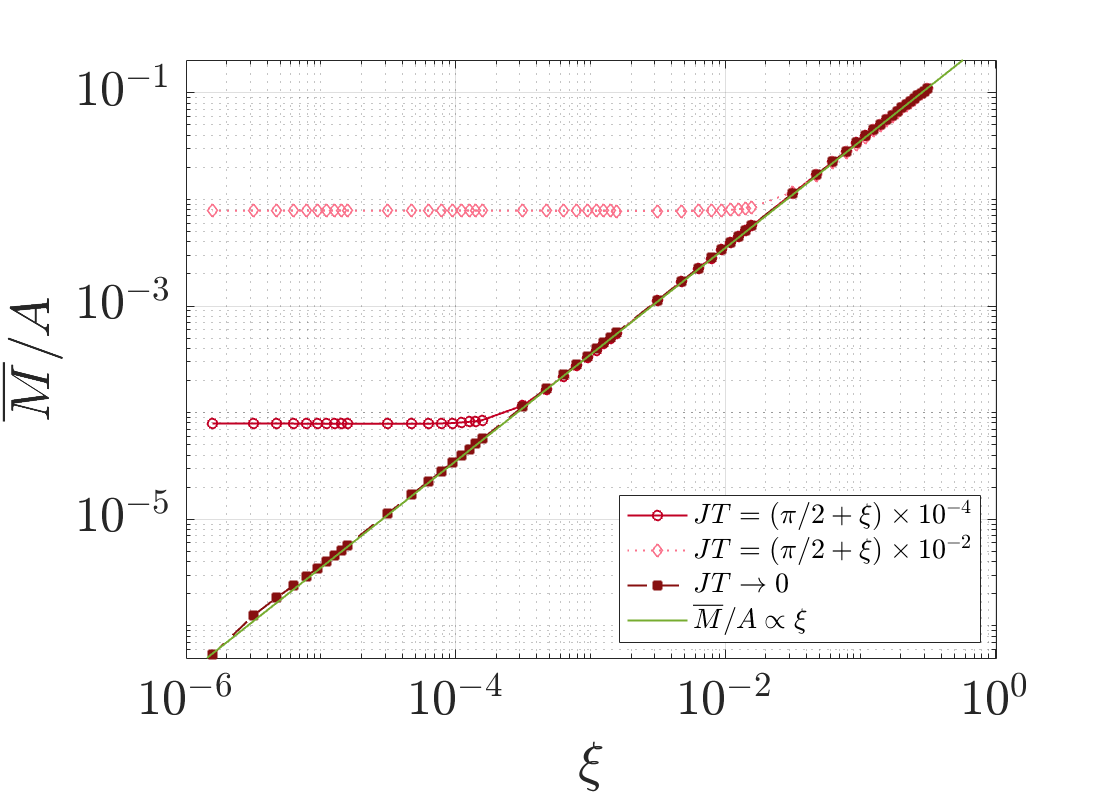}}}\hfill 
\caption{(a) $\overline{M}/A$ in units of $-e/\hbar T$ versus $\beta$ the inverse temperature for $\mu=\pi/2$, $JT/3 = 10^{-4} \times (\pi/2+\xi)$ and $\lambda J T/3 = \pi/2 + \xi$. Here we plot $\overline{M}/A$ for different $\xi=0.16,0.016,0.0016$ (diamonds, filled squares and circles, respectively). Solid-red line is $\propto \beta^3$. (b) $\overline{M}/A$ in units of $-e/\hbar T$ versus the detuning $\xi$, at a low temperature of $\beta\approx 1/1.6\times 10^{6} $ and $ \mu=\pi/2$. We choose  $\lambda JT/3=\pi/2+\xi$ and $JT/3 = (\pi/2+\xi) \times \left[0, 10^{-2}, 10^{-4}\right]$ corresponding respectively to filled squares, diamonds and circles. As $\beta \rightarrow \infty$ and $JT/3\rightarrow 0$, ${\overline M}/A \propto \xi$ for all $\xi$. On detuning with small but nonzero $JT/3 \ll 1$, the linear trend saturates as $\xi \rightarrow 0$.}
\label{Fig1a}
\end{figure*}

Fig.~\ref{Fig3c} shows the orbital magnetization for driven graphene slightly detuned from the flat band limit where a nonzero Berry curvature and hence a  nonzero orbital magnetization is induced. The orbital magnetization peaks at the band centers $\mu=\pm \pi/2$ and vanishes elsewhere, which is captured by our analytical expression Eq.~\eqref{Mex}. When the temperature increases, the width of the peaks broaden.

We now discuss the limit where 
the chemical potential lies in one of the bands 
so that $\mu = \pi/2$, and the temperature is high
$\beta \rightarrow 0$. In this limit, we obtain
\begin{eqnarray}
 \overline{M}/A 
&=& \xi^2 \frac{e}{ \hbar}\frac{\pi^3\beta^3}{12} + O(\xi^3). \label{highTlimit}
\end{eqnarray}

Fig.~\ref{Fig4a} demonstrates the orbital magnetization calculated numerically from Eq.~\eqref{Meq2}, and plotted with respect to $\beta$. Different detunings $\xi$ are chosen  with $JT/3= 10^{-4}\times(\pi/2+\xi)$ and $\lambda JT/3=\pi/2+\xi$. In addition, the chemical potential is chosen to lie in the band, $\mu=\pi/2$. Consistent with the high temperature limit of our analytical expression, we observe ${\overline M}/A \propto \beta^3$ for $\beta \rightarrow 0$. For low temperatures $\beta \rightarrow \infty$ the orbital magnetization becomes independent of $\beta$. In this regime of low temperatures, we also explore how the orbital magnetization depends on the detuning $\xi$, finding it to  be linear in $\xi$, Fig.~\ref{Fig4b}, in contrast to quadratic scaling in $\xi$ in the high temperature limit, Eq.~\eqref{highTlimit}. Fig.~\ref{Fig4b} also gives the scaling in $\xi$ in the flat band limit $JT\rightarrow 0, \lambda \rightarrow \infty$ and for small deviation from it realized by taking $JT \ll 1$ but nonzero. In the flat band limit, the linear scaling persists as $\xi \rightarrow 0$, however it eventually saturates to a nonzero value for $JT \ll 1$.

It is interesting to compare our analytic expression in Eq.~\eqref{Mex} to that obtained in Ref.~\cite{Gromov21} for a completely localized and disorder-free system. They found (setting $e=\hbar=1$) the orbital magnetization per unit area to be $  {\overline M}/A = T^{-1}/(e^{\alpha}+1)$, where $\alpha$ plays the role of a chemical potential $\alpha \equiv \beta \mu$. We note that Ref.~\cite{Gromov21} assumes a Floquet protocol that gives $U_F=I$ on a square lattice in the absence of a background gauge field. When the vector potential is introduced with minimal coupling, the Floquet unitary becomes diagonally distributed with phases determined by the flux picked up by the fermions. Hence the resulting orbital magnetization is built on the localized bulk states and the background gauge field. In contrast, in our calculation the Floquet unitary resulting after one period of driving does not lead to localized bulk bands, and hence it is not diagonally distributed, see Eqs.~\eqref{generalU} and~\eqref{UF1} (and also Appendix ~\ref{appC} for the Floquet unitary in Fock space). In this sense, Eq.~\eqref{Mex} contains both topological and non-topological contributions.

\section{Orbital Magnetization of the Haldane model: Static and Floquet}\label{HaldaneF}

\begin{figure*}
\centering{\includegraphics[width=0.33\textwidth]{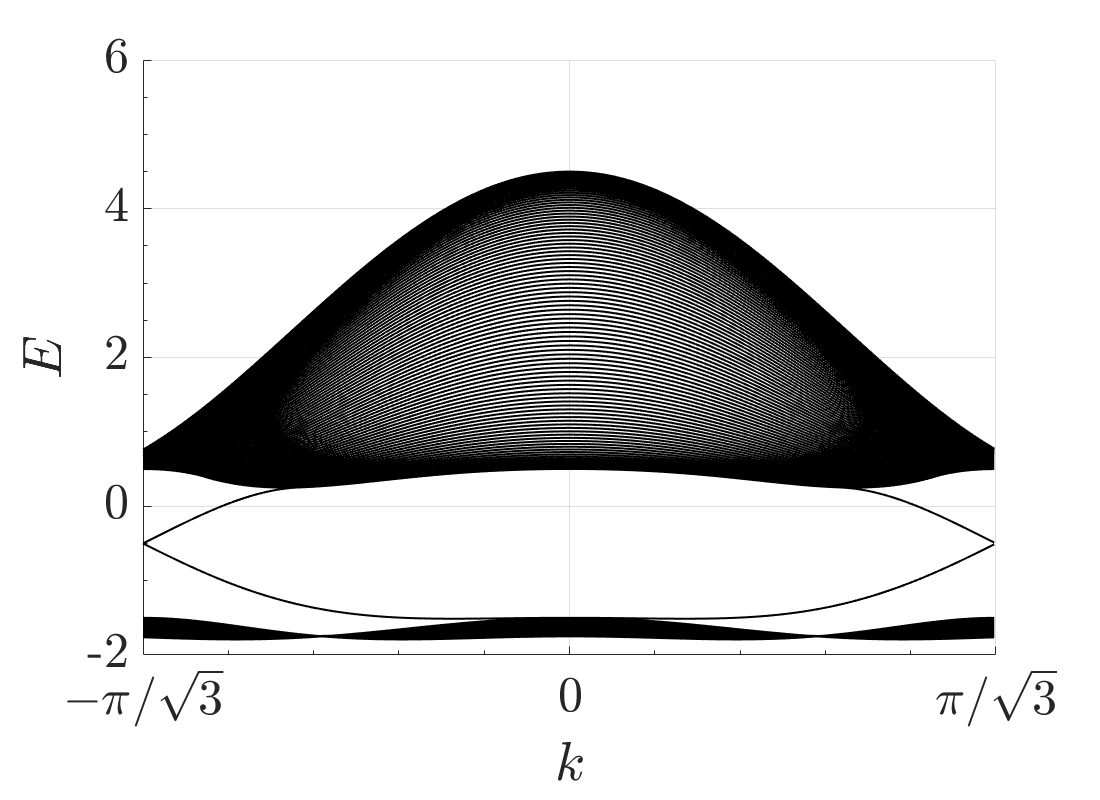}}\hfill \centering{\includegraphics[width=0.32\textwidth]{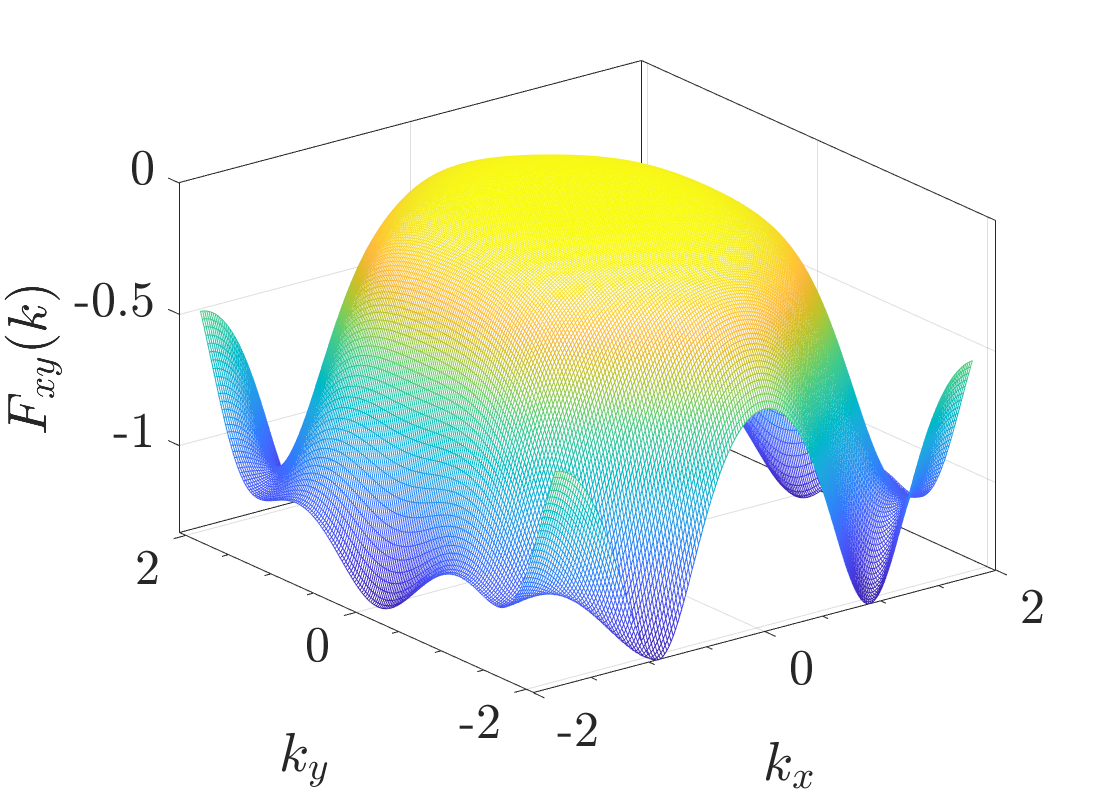}}
\centering{\includegraphics[width=0.32\textwidth]{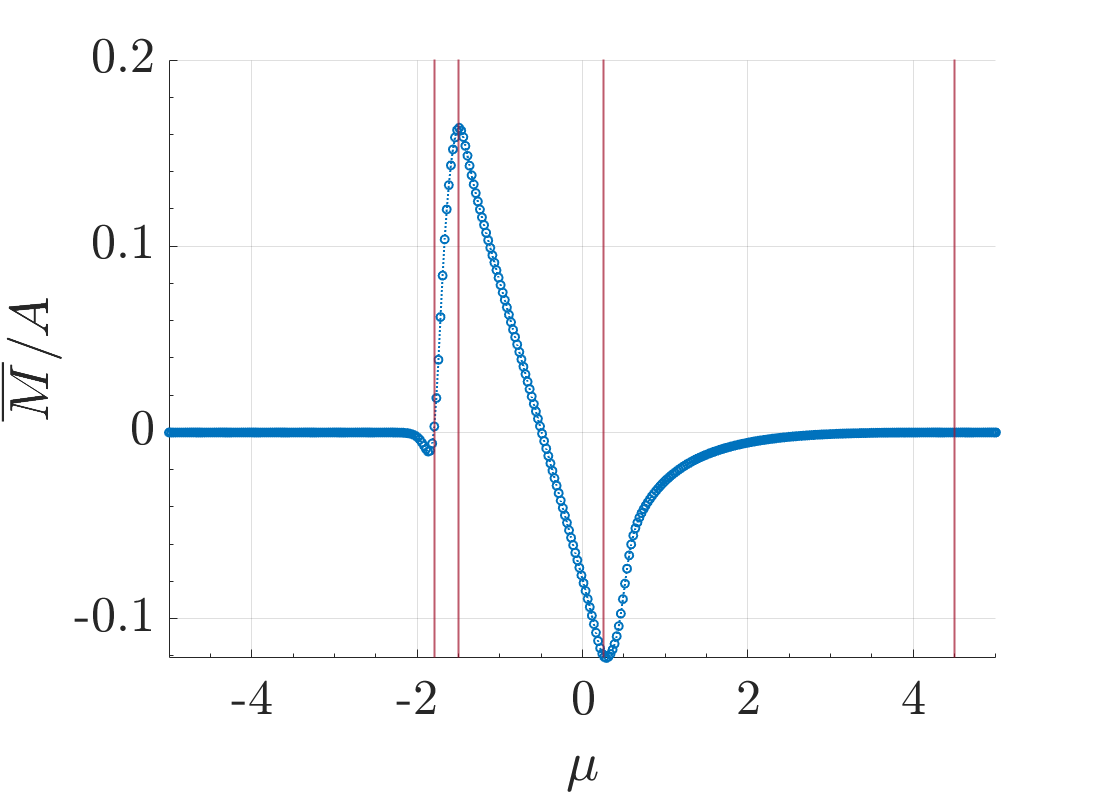}}
\caption{Results for the static Haldane model with $t_2=J/(12 \sqrt{3/43})$ and $\phi= \arccos(3\sqrt{3/43})$. (a) Spectrum on the cylinder with edge modes. (b) Berry curvature for the torus geometry. (c) Orbital magnetization per unit area on the torus, in units of $-e/\hbar T$ for temperature $\beta^{-1}= 0.05 \times J T/3$. Vertical red lines denote the band edges.}
\label{Fig4}
\end{figure*}

We now turn to the Haldane model Eq.~\eqref{HalF}. We first discuss the static case which corresponds to $\lambda=1$ in Eq.~\eqref{Jprofile}. It was shown \cite{Mudry11} that for
parameters where the flux $\phi= \arccos(3\sqrt{3/43})$ and ratio of the n.n.n.~to~n.n. hopping is $t_2=J/(12 \sqrt{3/43})$, the model in fact has a flat band. The Haldane model also breaks particle-hole symmetry for general flux. In particular, for the chosen parameters only one of the bands is flat. We first study the orbital magnetization  of the static model.

Fig.~\ref{Fig4} (a) shows the spectrum of the static model on a cylinder for the above-mentioned parameters. The flatness of
the lower band is apparent. Moreover, $C=1$, and this gives rise to a pair of chiral edge modes, one on each end of the cylinder. The Berry curvature of one of the bands is shown in Fig.~\ref{Fig4}(b). Despite the dispersion of the two bands being very different, the Berry curvature of one band is exactly negative to the Berry curvature of the other band.  
Fig.~\ref{Fig4}(c) shows the orbital magnetization. The temperature chosen ($\beta^{-1}=0.05 JT/3$) is small as compared to the hopping strength, thus when $\mu$ goes out of the band edges, the magnetization rapidly falls to zero. As the chemical potential traverses a band, the orbital magnetization peaks, with the sign of the
orbital magnetization being opposite in the two bands due to the opposite signs of the Berry curvature. Moreover, as the chemical potential traverses the gap between bands it changes linearly as follows $\overline{M}/A = -(e/\hbar)C \mu/2\pi$. 
The broken particle-hole symmetry is apparent in the non-zero value of the orbital magnetization when $\mu=0$,
i.e.,~at half-filling, and by the fact that 
the orbital magnetization is not perfectly anti-symmetric around $\mu=0$. 

\begin{figure*}
\subfloat[]{\label{Fig5a}\includegraphics[width=0.33\textwidth]{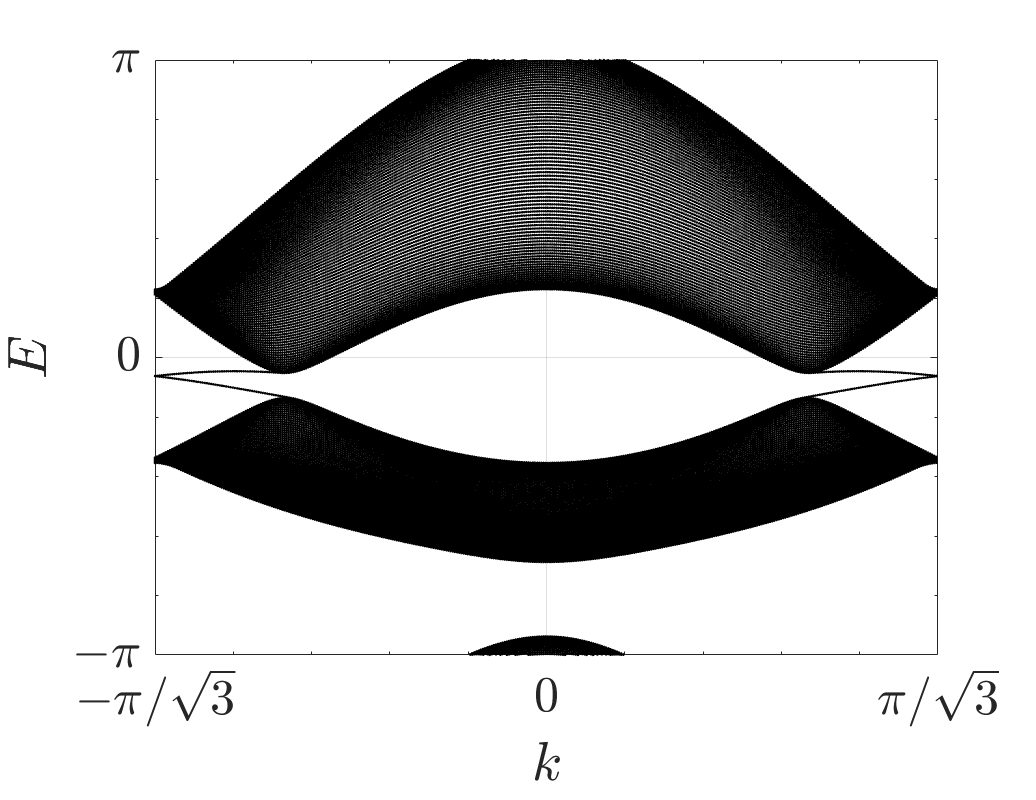}}\hfill 
\subfloat[]{\label{Fig5b}\includegraphics[width=0.33\textwidth]{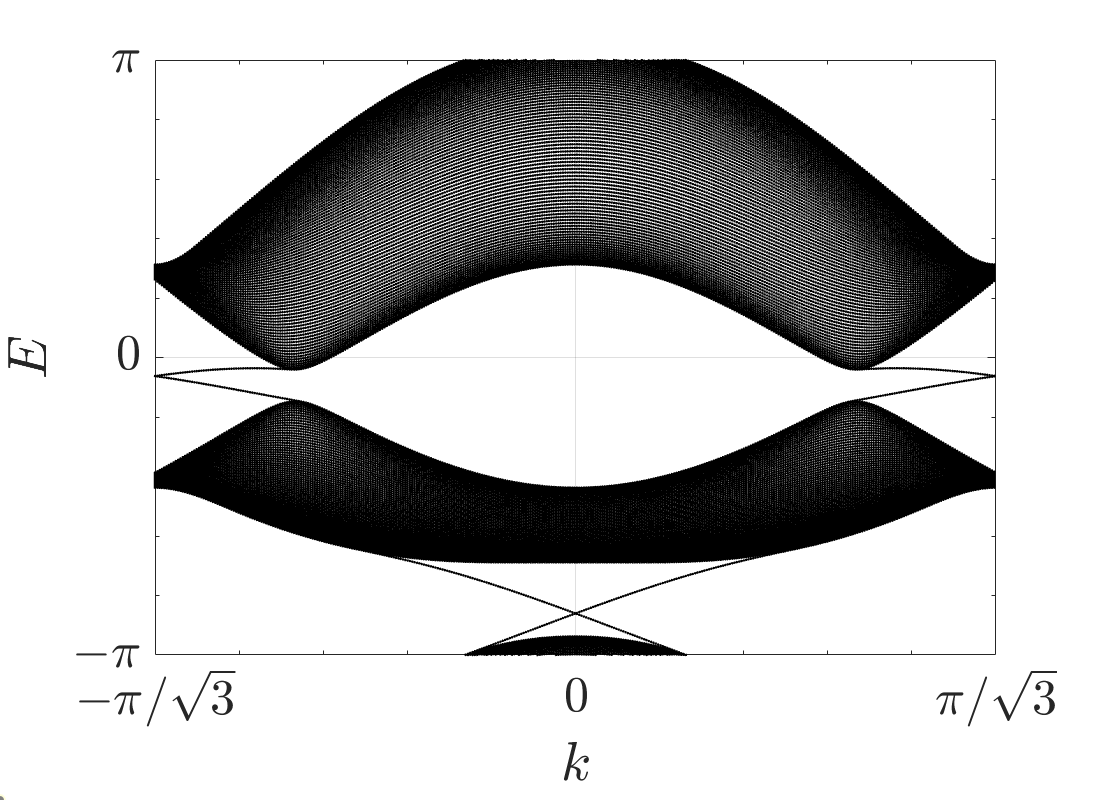}}\hfill 
\subfloat[]{\label{Fig5c}\includegraphics[width=0.33\textwidth]{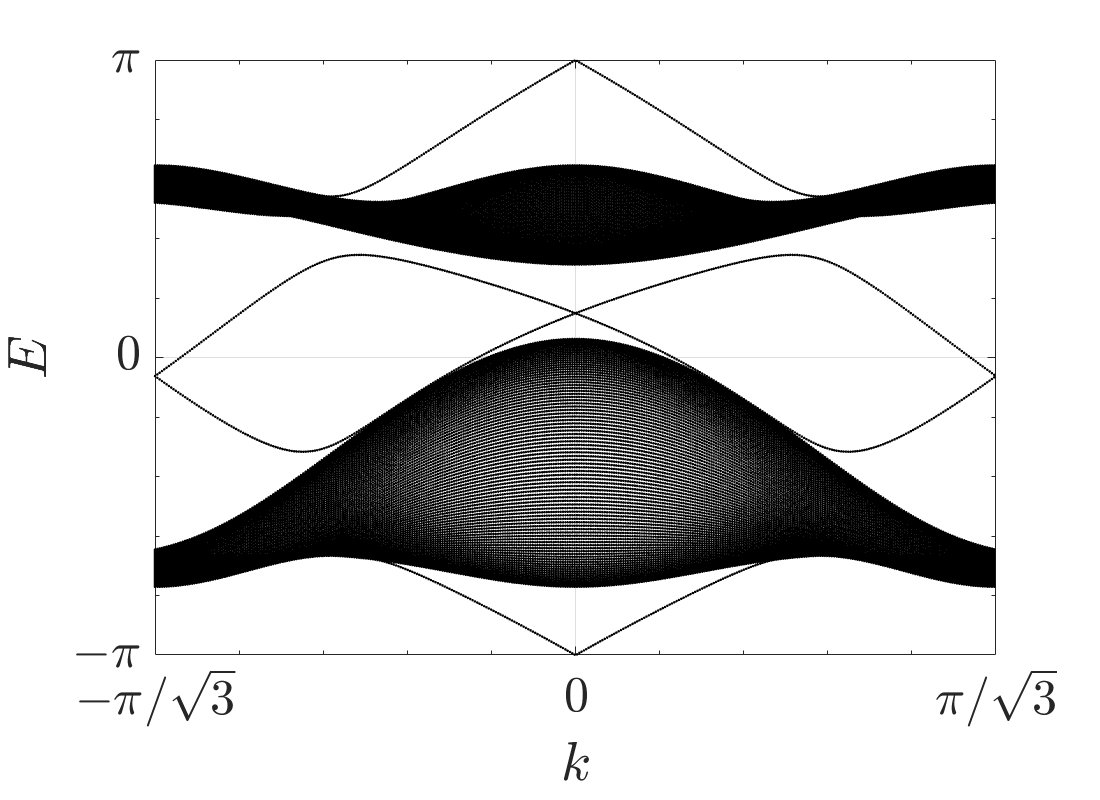}}\hfill 
\subfloat[]{\label{Fig5d}\includegraphics[width=0.33\textwidth]{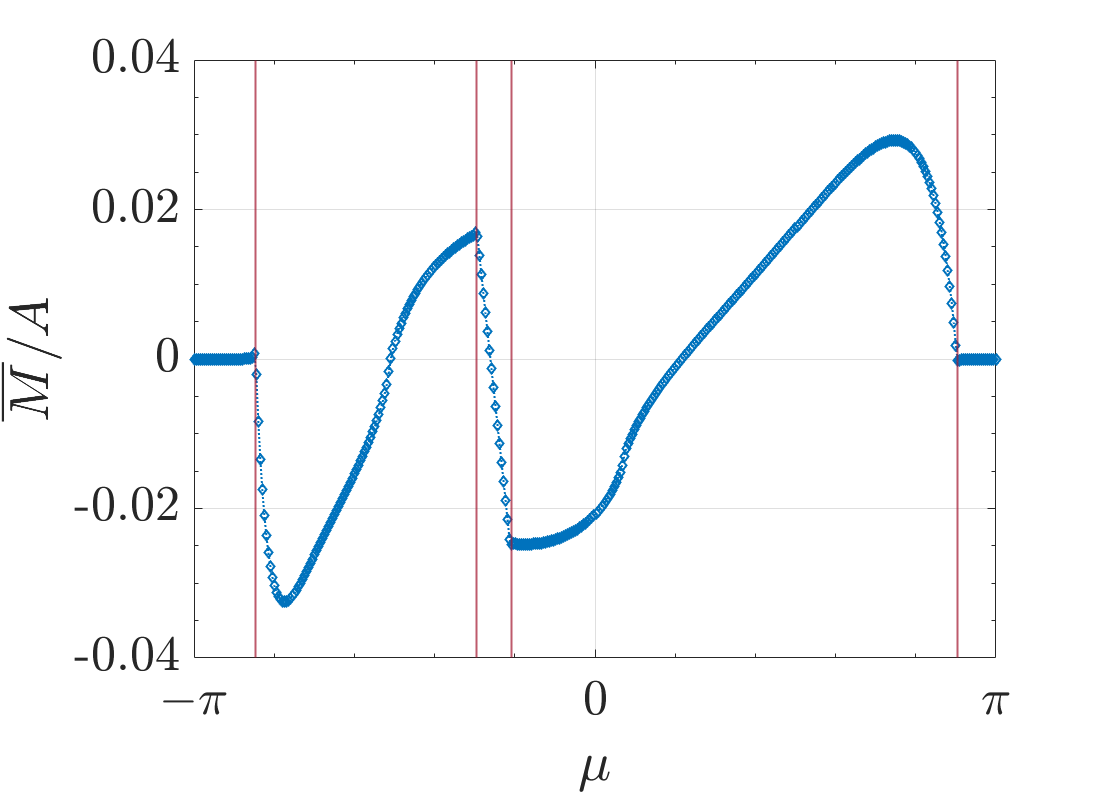}}\hfill
\subfloat[]{\label{Fig5e}\includegraphics[width=0.33\textwidth]{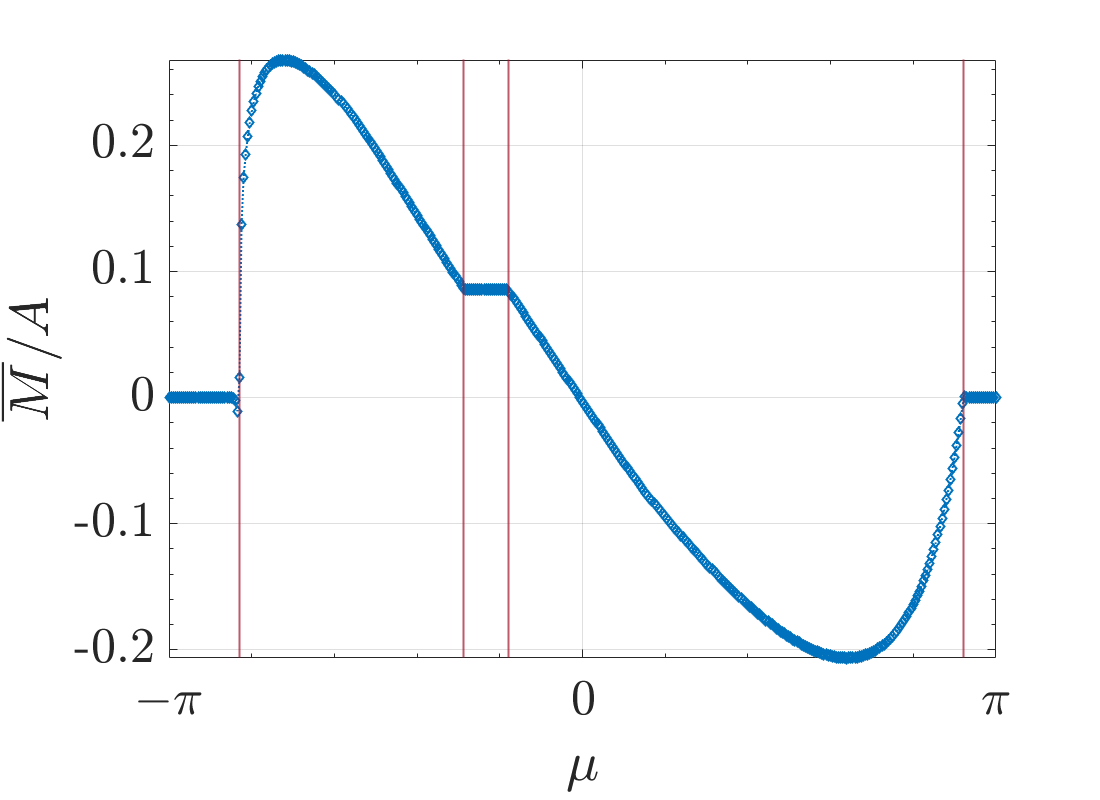}}\hfill 
\subfloat[]{\label{Fig5f}\includegraphics[width=0.33\textwidth]{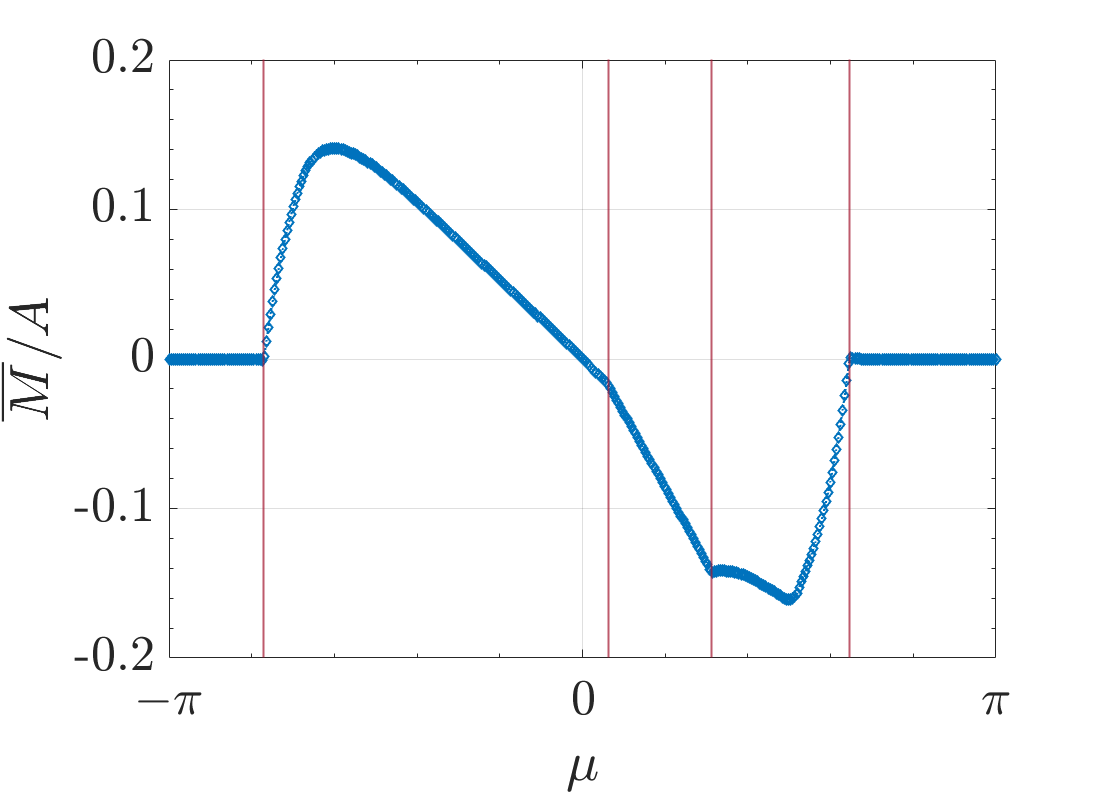}}\hfill 
\caption{ (a,b,c) Spectra on a cylinder for $N=100$. (d,e,f) Orbital magnetization per unit area in units of $-e/\hbar T$ for the model on the torus for temperature $\beta^{-1}= 0.05 \times J T/3$.  All figures are for $JT=\pi/8$, $t_2=J/(12 \sqrt{3/43})$ and $\phi= \arccos(3\sqrt{3/43})$ but different $\lambda$.  (a,d) $\lambda=5$ with $C=1$ where the edge modes are inherited from the static model. (b,e) $\lambda=7$  with $C=0$. This is an anomalous phase that shows two pairs of chiral edge modes. (c,f) $\lambda=15$ with $C=1$.  Note the shifts  in the quasi-energy in (d,e) by 
$-0.5$ and $-0.44$, respectively. This is needed for the proper calculation of the Berry curvature. See text for the details.}
\label{Fig5}
\end{figure*} 

We now Floquet drive the Haldane model. The protocol used is the same as that for graphene, Eq.~\eqref{Jprofile}, and with all other parameters held fixed in time. We choose  $\phi= \arccos(3\sqrt{3/43})$, and $t_2=J/(12 \sqrt{3/43})$, the same as in the static Haldane model. In addition, we set $JT=\pi/8$ while we vary the anisotropy parameter $\lambda$. 

Fig.~\ref{Fig5} shows results  for three different values of $\lambda$. In particular, $\lambda=5$ (a,d) gives a regular Chern insulator with $C=1$, and a corresponding pair of chiral edge modes. $\lambda=7$  (b,e) gives an anomalous Floquet phase with $C=0$, and two pairs of chiral edge modes. $\lambda=15$ again gives a regular Chern insulator with $C=1$, and a pair of chiral edge modes, but with these edge modes traversing the Floquet zone boundary. The Floquet zone center for this case hosts edge modes, but they are not topological edge modes as they do not start from one bulk band and terminate on another. Therefore the edge modes at the zone center may be removed without a gap closing. 

We note that the Floquet spectrum folds in Fig.~\ref{Fig5a} and~\ref{Fig5b} due to the broken particle-hole symmetry of the Haldane model. Accordingly, we observe that the edge modes do not appear exactly at $\epsilon=0$ and $\epsilon=\pm\pi$. In such cases one has to unfold the band to perform the Berry curvature and orbital magnetization calculations. We unfold by simply adding an overall energy shift to the Hamiltonian and hence shifting the Floquet spectrum. Due to the periodicity of the Floquet spectrum, the resulting bands still lie between $\left[-\pi,\pi\right]$, see red vertical lines in Figs.~\ref{Fig5d}-\ref{Fig5e}. 
Thus for low enough temperatures, we observe vanishing orbital magnetization outside the band-edges, and therefore at $\mu=\pm \pi$, Figs.~\ref{Fig5d}-\ref{Fig5e}.

Due to the broken particle-hole symmetry of the Haldane model, the orbital magnetization per unit area for these three cases
is non-zero at half-filling. In contrast, particle-hole symmetric systems such as Floquet driven graphene have zero orbital magnetization
at half-filling.  In addition, the orbital magnetization of Floquet driven Haldane model shows qualitatively the same behavior as that of the static Haldane model (Fig.~\ref{Fig4}) in that it vanishes outside the band edges, its variation with $\mu$ between the two bands is proportional to $C$, and it changes sign between the two bands.  Although the anomalous phase with two pairs of edge modes in Fig.~\ref{Fig5e} exhibits an orbital magnetization which is significantly larger in magnitude than the Chern insulator phase in Fig.~\ref{Fig5d} with only one pair of edge modes, let us emphasize that the bulk modes here are delocalized, and contribute to the orbital magnetization significantly. Therefore, it is not really possible to differentiate the effect of topological edge modes from bulk modes on the orbital magnetization.

\section{Conclusions} \label{Conclu}

We presented results for a Floquet system in two spatial dimensions, where the Floquet drive breaks an effective TRS. We explored regular Chern insulator phases as well as anomalous Floquet phases where the Chern number cannot fully characterize the edge modes of the system. In particular, we explored anomalous phases where the Chern number is zero and yet chiral edge modes exist in the system. 

We identified certain parameters for which the Floquet bands of an anomalous Floquet phase with $C=0$ are flat. We then derived expressions for the edge and bulk modes (Eq.~\eqref{generalU}), the
Berry curvature (Eq.~\eqref{Fxy0}), and the orbital magnetization (Eq.~\eqref{Mex}) to leading non-trivial order in the flatness of the bands (the latter quantified by $\xi$, c.f. Eq.~\eqref{limit2}). Since the Chern number is zero, the integral of the Berry curvature in Eq.~\eqref{Fxy0} over the momentum Brillouin zone vanishes. Nevertheless, the non-zero Berry curvature makes the orbital magnetization non-zero. We showed that the orbital magnetization is peaked when the chemical potential lies in the band. Furthermore, we explored the 
high temperature and low temperature properties of the orbital magnetization (c.f.,~Fig.~\ref{Fig1a}).

We also presented results for the orbital magnetization away from the flat band limit. In this context we explored the Floquet driven Haldane model where we showed that the broken particle-hole symmetry of the Haldane model enhances the orbital magnetization at half filling. In addition, 
this orbital magnetization was large even in the anomalous Floquet phase with $C=0$.

While Floquet driving is an active  area of study, linear response properties of these systems that go beyond Hall conductivity need to be further explored. In this paper we explored the orbital magnetization which is a linear response to a weak perturbing magnetic field. Other forms of linear-responses such as response of the system to weak strains, could help further uncover measurable consequences of Floquet induced topology. Studying the effect of interactions when the Floquet bands are flat, is also a natural direction of study. 

\emph{Acknowledgements:} This work was supported by the US Department of Energy, Office of Science, Basic Energy Sciences, under Award No.~DE-SC0010821 (AM). C.B.D. acknowledges support from an NSF grant for the Institute for Theoretical Atomic, Molecular, and Optical Physics at Harvard University and the Smithsonian Astrophysical Observatory.

%\bibliographystyle{apsrev4-2}
%\bibliography{orbmag}


\appendix
\section{Derivation of Eq.~\eqref{generalU}} \label{appA}
For the first site we have
\begin{align}
 & U_3^{\dagger} c_{1,k} U_3=i c_{2,k} e^{i k a \sqrt{3}/2},\\
 & U_2^{\dagger} c_{2,k} U_2 = i c_{1,k} e^{i k a \sqrt{3}/2},\\
  & U_1^{\dagger} c_{1,k} U_1 =c_{1,k}.
\end{align}
Thus,
\begin{align}
  U^{\dagger} c_{1,k} U = -c_{1,k} e^{i k a \sqrt{3}}.
\end{align}

Repeating for the next site,
\begin{align}
 & U_3^{\dagger} c_{2,k} U_3=i c_{1,k} e^{-i k a \sqrt{3}/2},\\
 & U_2^{\dagger} c_{1,k} U_2  = i c_{2,k} e^{-i k a \sqrt{3}/2},\\
  & U_1^{\dagger} c_{2,k} U_1 =i c_{3,k}.
\end{align}
Thus,
\begin{align}
  U^{\dagger}c_{2,k} U = -i c_{3,k} e^{-i k a \sqrt{3}}.
\end{align}

Repeating for the third site,
\begin{align}
 & U_3^{\dagger} c_{3,k} U_3=i c_{4,k} e^{i k a \sqrt{3}/2},\\
 & U_2^{\dagger} c_{4,k} U_2  = i c_{3,k} e^{i k a \sqrt{3}/2},\\
  & U_1^{\dagger} c_{3,k} U_1 =i c_{2,k}.
\end{align}
Thus,
\begin{align}
  U^{\dagger}c_{3,k} U = -i c_{2,k} e^{i k a \sqrt{3}}.
\end{align}
Therefore the matrix $\tilde{U}$ for a 8-site system has the form
of Eq.~\eqref{generalU}.

\section{Derivation of Eq.~\eqref{Mex}} \label{appB}

We consider the limit in Eq.~\eqref{limit2}. For this case,
denoting 
\begin{align}
U_F(\boldsymbol{k}) &= \epsilon(k)+\sum_i d_i(\boldsymbol{k}) \sigma_i,\\
    \lambda JT/3 &= \pi/2 + \xi,
\end{align}
where $\epsilon(\boldsymbol{k})$ is the coefficient of the identity matrix and $d_1,d_2,d_3$ are the coefficients of the Pauli matrices. We find these to be
\begin{widetext}
\begin{eqnarray}
\epsilon(\boldsymbol{k}) &=& \cos^3(J\lambda T/3) - \cos(J\lambda T/3) \sin^2(J\lambda T/3) \left[\cos(\sqrt{3}k_x)+2\cos\left(\frac{3k_y}{2}\right)\cos\left(\frac{\sqrt{3}k_x}{2}\right)\right],\\
d_1(\boldsymbol{k}) &=& -i \sin(J\lambda T/3) \cos^2(J\lambda T/3) \left[\cos(k_y)+2\cos\left(\frac{k_y}{2}\right)\cos\left(\frac{\sqrt{3}k_x}{2}\right)\right]+ i\sin^3(J\lambda T/3) \cos(\sqrt{3}k_x+k_y)
,\\
d_2(\boldsymbol{k}) &=& -i \sin(J\lambda T/3) \cos^2(J\lambda T/3) \left[\sin(k_y)-2\sin\left(\frac{k_y}{2}\right)\cos\left(\frac{\sqrt{3}k_x}{2}\right)\right]+ i\sin^3(J\lambda T/3) \sin(\sqrt{3}k_x+k_y),\\
d_3(\boldsymbol{k}) &=& i \cos(J\lambda T/3) \sin^2(J\lambda T/3) \left[\sin(\sqrt{3}k_x)-2\sin\left(\frac{3k_y}{2}\right)\cos\left(\frac{\sqrt{3}k_x}{2}\right)\right],
\end{eqnarray}
with the eigenvalues and eigenfunctions being
\begin{eqnarray}
E_{\pm} &=& \epsilon(\boldsymbol{k}) \pm \sqrt{d_1^2(\boldsymbol{k})+d_2^2(\boldsymbol{k})+d_3^2(\boldsymbol{k})} = \epsilon(\boldsymbol{k})\pm d(\boldsymbol{k}), \\
\psi_{\pm} &=& \frac{1}{\sqrt{2d(\boldsymbol{k})(d(\boldsymbol{k}) \pm d_3(\boldsymbol{k}))}}\left( \begin{array}{c}
d_3(\boldsymbol{k})\pm d(\boldsymbol{k}) \\
d_1(\boldsymbol{k})-id_2(\boldsymbol{k})   
\end{array} \right).
\end{eqnarray}
We will use the following formula for the Berry curvature
\begin{eqnarray}
F_{xy}(\boldsymbol{k}) = \frac{1}{2d^3(\boldsymbol{k})} \epsilon_{abc} d_a \partial_x d_b \partial_y d_c. 
\end{eqnarray}
It is helpful to note that $\epsilon(\boldsymbol{k}) = -\xi \epsilon_1(\boldsymbol{k}) + O(\xi^3)$, where $\epsilon_1(\boldsymbol{k})$ is given in Eq.~\eqref{eps1}. In addition, 
$d_1(\boldsymbol{k}) = i \cos(\sqrt{3}k_x+k_y) + O(\xi^2)$, $d_2(\boldsymbol{k}) = i \sin(\sqrt{3}k_x+k_y) + O(\xi^2)$, $d_3(\boldsymbol{k}) = O(\xi)$. Thus $d^2(\boldsymbol{k}) = -1 + O(\xi^2)$.
Thus, to the lowest order in $\xi$, the Berry curvature is given by Eq.~\eqref{Fxy0}. It is straightforward to see that the integration of the
Berry curvature over the Brillouin zone of graphene is zero, corresponding to a Chern number of $C=0$.

We now consider the following term, 
\begin{eqnarray}
\epsilon_{u\boldsymbol{k}}+\epsilon_{d\boldsymbol{k}} &=& i\log(E_{-})+i\log(E_{+}) = i\log\left(E_{-}E_{+}\right)
= i \log\left[\epsilon^2(\boldsymbol{k})-d^2(\boldsymbol{k}) \right].
\end{eqnarray}
It is then straightforward to see that
\begin{eqnarray}
\epsilon_{d\boldsymbol{k}}+\epsilon_{u\boldsymbol{k}} &=& 0, 
\end{eqnarray}
due to preserved particle-hole symmetry. We also need the difference between the Fermi functions
\begin{eqnarray}
f_{d\boldsymbol{k}} - f_{u\boldsymbol{k}} &=& \frac{1}{1+\exp[\beta(\log E_+^i-\mu)]} - \frac{1}{1+\exp[\beta(\log E_-^i-\mu)]} \notag \\
&=& \frac{1}{1+ E_+^{i\beta} e^{-\beta \mu}} - \frac{1}{1+ E_-^{i\beta} e^{-\beta \mu}} = e^{-\beta\mu}\frac{E_-^{i\beta}  - E_+^{i\beta}}{1+ E_+^{i\beta}e^{-\beta \mu}+E_-^{i\beta}e^{-\beta \mu}+ (E_+E_-)^{i\beta}e^{-2\beta \mu} }.
\end{eqnarray}
Expanding $E_{\pm}$ in $\xi$, one obtains Eq.~\eqref{fdiff}.

We also need to evaluate 
$(\epsilon_{u\boldsymbol{k}}-\epsilon_{d\boldsymbol{k}})\sum_{n=d,u}f'_{n\boldsymbol{k}}(\epsilon_{n\boldsymbol{k}}-\mu)$ to $O(\xi)$. 
For this, we  use that
\begin{align}
    \epsilon_{n\boldsymbol{k}} &=\pm \frac{\pi}{2} \pm  
    \xi \epsilon_1(\boldsymbol{k}) \pm O(\xi^2),\label{B14}
\end{align}
so that $\epsilon_{u\boldsymbol{k}}-\epsilon_{d\boldsymbol{k}} = \pi + 2\xi \epsilon_1(\boldsymbol{k}) +2 O(\xi^2)$. Thus we need to evaluate
$\sum_{n=d,u}f'_{n\boldsymbol{k}}(\epsilon_{n\boldsymbol{k}}-\mu)$ to $O(\xi)$.
We write
\begin{align}
    \sum_{n=d,u}f'_{n\boldsymbol{k}}(\epsilon_{n\boldsymbol{k}}-\mu) = -\sum_{n=d,u}\beta(\epsilon_{n\boldsymbol{k}}-\mu)\frac{e^{\beta(\epsilon_{n\boldsymbol{k}}-\mu)}}{(1 + e^{\beta(\epsilon_{n\boldsymbol{k}}-\mu)})^2}.
\end{align}
Substituting Eq.~\eqref{B14} in the above expression and expanding in $\xi$,
we obtain Eq.~\eqref{Eq19} in the main text.

%\begin{align}
%&-\sum_{n=d,u}f'_{nk}\beta(\epsilon_{nk}-\mu) =\frac{1}{2} \beta  e^{\beta  \mu +\frac{\pi  \beta }{2}} \left(\frac{\pi -2 \mu }{\left(e^{\beta  \mu }+e^{\frac{\pi  \beta }{2}}\right)^2}-\frac{2 \mu +\pi }{\left(e^{\frac{1}{2} \beta  (2 \mu +\pi )}+1\right)^2}\right)\nonumber\\
%&+\frac{1}{8} \beta  \epsilon_1(k) \xi \biggl[\biggl\{\beta  (\pi -2 \mu ) \tanh \left(\frac{1}{4} \beta  (\pi -2 \mu )\right)-2\biggr\} \text{sech}^2\left(\frac{1}{4} \beta  (\pi -2 \mu )\right)\nonumber\\
%&+\biggl\{\beta  (2 \mu +\pi ) \tanh \left(\frac{1}{4} \beta  (2 \mu +\pi )\right)-2\biggr\} \text{sech}^2\left(\frac{1}{4} \beta  (2 \mu +\pi )\right)\biggr]+O\left(\xi^2\right).
%\end{align}

\section{Analytic expression for the Floquet unitary on a cylinder in Fock space} \label{appC}

Here we derive an analytical expression for the Floquet unitary on a cylinder in the limiting case of $\xi=0$. The entire Floquet unitary reads
\begin{eqnarray}
U=U_3 U_2 U_1. \notag
\end{eqnarray}
When the unitaries at each step are expanded for $\lambda \rightarrow \infty, JT\rightarrow 0$ but $\lambda J T$ finite, one arrives at the following expressions
\begin{align}
  U_1 &= \prod_{j} \left[1 + \bigg\lbrace\cos \left(\frac{J \lambda T}{3}\right) -1\bigg\rbrace (n_{2j,k}-n_{2j+ 1,k})^2 +  i\sin\left(\frac{J \lambda T}{3}\right)\left(c_{2j,k}^{\dagger}c_{2j+ 1,k} + h.c.\right)\right], \notag \\
   U_2 &= \prod_{j} \left[1 + \bigg\lbrace\cos \left(\frac{J \lambda T}{3}\right) -1\bigg\rbrace (n_{2j,k}-n_{2j- 1,k})^2 +  i\sin\left(\frac{J \lambda T}{3}\right)\left(c_{2j,k}^{\dagger}c_{2j- 1,k}e^{i k a \sqrt{3}/2} + h.c.\right)\right], \notag \\
  U_3 &= \prod_{j} \left[1 + \bigg\lbrace\cos \left(\frac{J \lambda T}{3}\right) -1\bigg\rbrace (n_{2j,k}-n_{2j- 1,k})^2 +  i\sin\left(\frac{J \lambda T}{3}\right)\left(c_{2j,k}^{\dagger}c_{2j- 1,k}e^{-i k a \sqrt{3}/2} + h.c.\right)\right].\notag
\end{align}
where $n_{j,k}=c_{j,k}^{\dagger} c_{j,k}$. Assuming $JT\lambda/3=\pi/2$ as chosen in the main text, the entire unitary can be computed 
\begin{eqnarray}
U &=& \prod_{j}  \left[ \bigg\lbrace 1 - (n_{2j,k}-n_{2j- 1,k})^2 \bigg\rbrace +  i (e^{-i k a \sqrt{3}/2}c_{2j,k}^{\dagger}c_{2j-1,k}+h.c.) \right] \notag \\
&\times & \left[\bigg\lbrace 1 - (n_{2j,k}-n_{2j- 1,k})^2 \bigg\rbrace +  i (e^{i k a \sqrt{3}/2}c_{2j}^{\dagger}c_{2j-1}+h.c.)\right] \notag \\
& \times & \left[\bigg\lbrace 1 - (n_{2j,k}-n_{2j+ 1,k})^2 \bigg\rbrace + i \left(c_{2j,k}^{\dagger}c_{2j+ 1,k} + h.c.\right)\right].  \label{Ucylinder}
\end{eqnarray}
The first two lines above corresponds to the multiplication $U_3 U_2 = \prod_j N_{j,k}$, where $N_{j,k}$ is diagonal in the occupation number basis and is given by
\begin{eqnarray}
N_{j,k} = \left[1 -  n_{2j,k} (1+e^{-i k a \sqrt{3}/2}) - n_{2j-1,k} (1+ e^{i k a \sqrt{3}/2}) + 2n_{2j-1,k} n_{2j,k} \bigg\lbrace \cos\left(\frac{\sqrt{3}a k}{2}\right)+1\bigg \rbrace\right].\label{Nj}
\end{eqnarray}
Then substituting Eq.~\eqref{Nj} into Eq.~\eqref{Ucylinder}, we obtain
\begin{eqnarray}
U&=&\prod_j \biggl[N_{j,k} \lbrace 1 - \left( n_{2j,k}-n_{2j+ 1,k}\right)^2\rbrace + iN_{j,k} \left(c_{2j,k}^{\dagger}c_{2j+ 1,k} + h.c.\right)\biggr]. 
\end{eqnarray}
Then elements of the Floquet unitary that are diagonal in the number basis become
\begin{eqnarray}
N_{j,k} \lbrace 1 - \left( n_{2j,k}-n_{2j+ 1,k}\right)^2\rbrace  &=& 1 - \left( n_{2j,k}-n_{2j+ 1,k}\right)^2 - n_{2j,k} n_{2j+1,k} \left(1+e^{-i k a \sqrt{3}/2}\right) + \bigg[ n_{2j-1,k}(n_{2j,k}-1) \notag \\
&+& (-1)^{n_{2j,k}} n_{2j-1,k}n_{2j+1,k}  \bigg]\left(1+ e^{i k a \sqrt{3}/2}\right) + 2 n_{2j,k} \bigg[ n_{2j-1,k}-n_{2j+1,k} \notag \\
&+& n_{2j-1,k}n_{2j+1,k} \bigg] \bigg[\cos\left(\frac{\sqrt{3}a k}{2}\right) + 1 \bigg].
\end{eqnarray}
One can see that the first off-diagonal elements (i.e., n.n. hopping terms) of the Floquet unitary are nonzero,
\begin{eqnarray}
iN_{j,k} \left(c_{2j,k}^{\dagger}c_{2j+ 1,k} + h.c.\right) &=& i \bigg [ 1 - n_{2j-1,k} \left(1+ e^{i k a \sqrt{3}/2}\right) \bigg] \left(c_{2j,k}^{\dagger}c_{2j+ 1,k} + h.c.\right) \notag \\
&+& i \bigg [ -  \left(1+ e^{-i k a \sqrt{3}/2}\right) +2 n_{2j-1,k} \bigg(\cos\left(\frac{\sqrt{3}a k}{2}\right) + 1 \bigg) \bigg] c_{2j,k}^{\dagger}c_{2j+ 1,k}.
\end{eqnarray}
 The above expressions show that after one Floquet cycle, $U$ is not diagonal but contains hopping terms, indicating that the  bulk bands are delocalized. This is in contrast to the driving scheme discussed in Ref.~\cite{Gromov21} where
$U$ was completely diagonal with the fermion returning back to its starting point at the end of the Floquet cycle.

\end{widetext}

\end{document}